\documentclass{jfm}

\usepackage{hyperref}
\usepackage[usenames, dvipsnames]{xcolor}
\hypersetup{colorlinks=true,linkcolor=MidnightBlue,citecolor=MidnightBlue}

\usepackage{graphicx}

\usepackage[utf8]{inputenc}
\usepackage{layouts}
\usepackage{bbm}
\usepackage{bigints}
\usepackage{multirow}
\usepackage{overpic}
\usepackage{tcolorbox}
\usepackage{todonotes}\presetkeys{todonotes}{inline,backgroundcolor=yellow,bordercolor=yellow}{}

\usepackage{enumitem}

\newcommand{\pfrac}[2]{\frac{\partial #1}{\partial #2}}
\newcommand{\pnfrac}[3]{\frac{\partial^{#1} #2}{\partial #3^{#1}}}
\newcommand{\Dfrac}[2]{\frac{\mathrm{d} #1}{\mathrm{d} #2}}
\newcommand{\Dnfrac}[3]{\frac{\mathrm{d}^{#1} #2}{\mathrm{d} #3^{#1}}}

\newcommand\upd{\mathrm{d}}
\newcommand*{\ep}{\epsilon}

\newcommand*{\px}{P}

\newcommand{\fig}[1]{figure~\ref{fgr:#1}}

\shorttitle{Shear-induced instabilities of flows through submerged vegetation }
\shortauthor{C. Y. H. Wong, P. H. Trinh and S. J. Chapman}

\title{Shear-induced instabilities of flows through submerged vegetation }

\author{Clint Y. H. Wong\aff{1}
	\corresp{\email{clint.wong@maths.ox.ac.uk}},
	Philippe H. Trinh\aff{2}
	\and S. Jonathan Chapman\aff{1}}

\affiliation{\aff{1}Oxford Centre for Industrial and Applied Mathematics, 
	Mathematical Institute, \\ University of Oxford, Oxford OX2 6GG, UK
	\aff{2}Department of Mathematical Sciences, University of Bath, Bath BA2 7AY, UK}
\begin{document}


\maketitle
	
\begin{abstract}
We consider the instabilities of flows through a submerged canopy, and show how the full governing equations of the fluid-structure interactions can be reduced to a compact framework that captures many key features of vegetative flow. By modelling the canopy as a collection of homogeneous elastic beams, we predict the steady configuration of the plants in response to a unidirectional flow. This treatment couples the beam equations in the canopy to the fluid momentum equations. Our linear stability analysis suggests new insights into the development of instabilities at the surface of the vegetative region. In particular, we show that shear at the top of the canopy is a dominant factor in determining the onset of instabilities known as \emph{monami}. Based on numerical and asymptotic analysis of the generalised eigenvalue problem, the system is shown to be stable if the canopy is sufficiently sparse or if the plants are sufficiently flexible.


\end{abstract}


	
	
\section{Introduction}
\label{sec:1_intro}
The study of fluid-structure interactions with vegetation has a wide range of industrial and environmental applications, including flood control, environmental conservation, and energy production. However, there are a number of challenges in modelling such flows, in particular due to the fact that the vegetation both affects and is affected by the flow. The focus of this work is to develop compact mathematical models that describe flows through submerged vegetated regions and their resultant instabilities. In particular, we predict the critical parameters in different regimes for instabilities which resemble \textit{monami}---the synchronous waving of vegetation. 

\begin{table}
	\centering
	\caption{Modelling approaches of a selection of previous work on flow through a fully-submerged canopy or a single plant. The \textquoteleft Coupling\textquoteright~category states whether the stability analysis takes into account of both flow perturbations and plant reconfiguration.}
	\scriptsize
	\vspace{0.2cm}
	\begin{tabular}{lcccc}
		& Flow    & Plant model  & Stability analysis &Coupling\\ \hline
		\citet{Alben2002}       & solved  & elastic strip          & no &    \\
		\citet{Ghisalberti2004} & solved  & rigid cylinder        & no &          \\
		\citet{Poggi2004a}    & solved  & rigid cylinder        & no   & \\
		\citet{Dupont2010}          & solved & mechanical oscillator & yes & yes       \\
		\citet{Luhar2011}       & imposed    & elastic strip         & no &    \\
		\citet{Luminari2016}    & solved  & rigid cylinder        & yes   & no \\
		\citet{Singh2016}       & solved  & rigid cylinder        & yes  & no \\
		\citet{Zampogna2016}    & solved  & rigid cylinders    & yes   &       no\\
		& solved  & rigid porous medium    & yes   &       no\\
		
		\citet{Sharma2017}      & solved  & dynamic cluster & yes  & yes    \\
		& solved  & rigid porous medium        & yes    & no  \\
		\citet{Leclercq2018}    & imposed & elastic beam          & yes & no \\
		This work       & solved        & elastic beam                       & yes    & yes 
	\end{tabular}
	\label{table:literature}
\end{table}

\subsection{Flow through aquatic vegetation}
\label{ssec:intro_flow}
\noindent Climate change is increasing the frequency and severity of hydrological disasters. With the rise of more challenging flow management problems, there is an increasing demand for new solutions. While there are many infrastructural solutions on flow management, there is also an emerging interest in utilising aquatic vegetation, as it is part of the natural habitat helping to sustain our ecosystems. Above all, compared to artificial measures, aquatic vegetation has the promising abilities of adapting to the local environment and self-repairing after destructive events \citep{Morris2018}. On the other hand, flows through vegetation are challenging to model due to interactions between the two. As such, the efficiency of aquatic vegetation in protecting coastal regions and the physical mechanisms involved are yet to be fully understood mathematically \citep{Marion2014}.  

Flows through aquatic vegetation are an interesting class of flows to study due to the geometric and mechanical properties of aquatic vegetation. For plants to avoid mechanical failure or being uprooted, they have evolved to be typically flexible and streamlined so that they can passively reconfigure to reduce the fluid load \citep{Vogel1994}. In addition to their deformable nature, they can also have complex geometries. Their components can have length scales that differ by multiple orders of magnitude. Finally, distinct from terrestrial flows, submerged vegetation has a typical height comparable in magnitude to the water depth in order to photosynthesise \citep{Marion2014}. As a result, a significant proportion of the flow is obstructed by the \textit{canopy}---a community of vegetation.  

There are numerous modelling challenges in capturing the macro-scale and micro-scale properties of the system, and this primarily relates to the feedback mechanism between flow and vegetation. In a complete dynamic model, the fluid will apply a load on each vegetative structure, which causes a resultant deformation and this, in turn, must affect the flow. Thus in general, it appears that the fluid must be solved simultaneously with the configurations of each structure, and it is not clear what reductions can be applied. These challenges associated with modelling flow through vegetation have demanded sophisticated studies; see \emph{e.g.}~experimental works by 
\cite{Dunn1996,Ghisalberti2004,Hu2014,Mandel2019} and numerical works by \cite{Mattis2013,Sundin2018,Sharma2019}.

The question that we seek to answer in this work is whether there exist simpler mathematical models that are able to capture a number of the key physical features obtained in more complicated formulations or experiments. In regards to the development of a simplified model, the vast majority of previous work has fallen in two categories: either models of flow over a specified set of rigid obstacles [see \emph{e.g.}~\citet{Ghisalberti2004}]; or models where deformation can occur, but only under a known imposed flow [see \emph{e.g.}~\citet{Luhar2011}]. There have been fewer models that emphasise the coupling between deformation and flow. We highlight, in particular, the work of \citet{Alben2002} on flow past a single elastic strip in 2D and \citet{Dupont2010} on flow past an array of rigid and straight elements that are free to tilt. A summary of previous work, and their key features is presented in table~\ref{table:literature}.

\subsection{Instabilities in aquatic vegetation}

\begin{figure}
	\centering
	\includegraphics{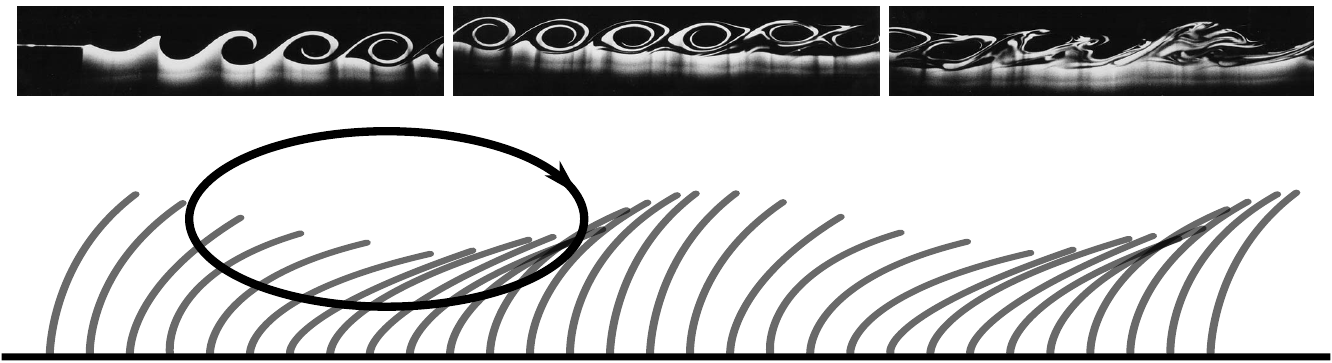}
	\caption{Schematic diagram of monami---the synchronous oscillation of plants. The grey obstacles indicate aquatic vegetation and the black arrow indicates a rolling vortex. The three insets are snapshots of Kelvin-Helmholtz instability developing along a channel. Two flows with different velocities meet upstream (left) and mix as they propagate downstream (right). The photos of the laboratory demonstration are extracted from \citet{Cushman-Roisin2005}.}
	\label{fgr:kh}
\end{figure}

\noindent Part of the emerging interest in instabilities of flow through a canopy is sparked by a fascinating phenomenon known as \emph{monami} (see the schematic in \fig{kh})---the progressive, synchronous oscillation of aquatic vegetation \citep{Nepf2012}. \textit{Honami}, its counterpart in terrestrial flows is readily observable in daily life when wind blows across a patch of grass or a crop field \citep{DeLangre2008}. Only recently has this phenomenon been explained by \citet{Raupach1996}, pointing out that such instabilities arise due to a shearing mechanism that resembles the Kelvin-Helmholtz instability on free mixing layers at the top of the canopy. Once the shear exceeds a threshold, waves develop in the flow which evolve into vortices over time (cf. insets of \fig{kh}). The argument that such instabilities are distinct from boundary layer instabilities is supported by comparing statistics of turbulent kinetic energy in experiments \citep{Finnigan2000, Poggi2004a}. 

Since this explanation has been accepted, many studies have attempted to explain and predict monami mathematically by analysing the stability of steady states in their respective model (cf. Table~\ref{table:literature}). In particular, \citet{Singh2016} established the dependence between viscous effects and flow instabilities by analysing flows through an array of rigid beams. 


\subsection{Objectives of this paper}
\noindent This article is dedicated to analysing the mechanical aspects of flow through submerged vegetation and its resultant instabilities. We develop a coupled model for the fluid flow and the mechanical deformation of the canopy, where the plants are allowed to have large angles of deflection. Using this model, which also accounts for viscous effects, we assess criteria and mechanisms for the onset of instability. Furthermore, determine for which flow regimes de-coupling between flow and vegetation is a reasonable approximation, which is currently unclear. 

The structure of this article is as follows. In \S\ref{sec:2_model}, we derive a mathematical model which couples the dynamics of the flow with the reconfiguration of the canopy, modelled as an array of elastic beams. Using this coupled model, we analyse flows which are steady and unidirectional in \S\ref{sec:3_uni}. In \S\S\ref{sec:4_LSA}--\ref{sec:6_crit}, we assess the temporal stability of such steady configurations. The analysis attempts to predict the critical parameters for instabilities which resemble monami and highlight how these parameters differ when the beams are rigid and vertical. We summarise our findings in \S\ref{sec:7_con} and discuss limitations and future work in \S\ref{sec:8_discuss}.

\section{Mathematical model}
\label{sec:2_model}
\begin{figure}
	\centerline{
	\begin{overpic}[width=0.7\textwidth]{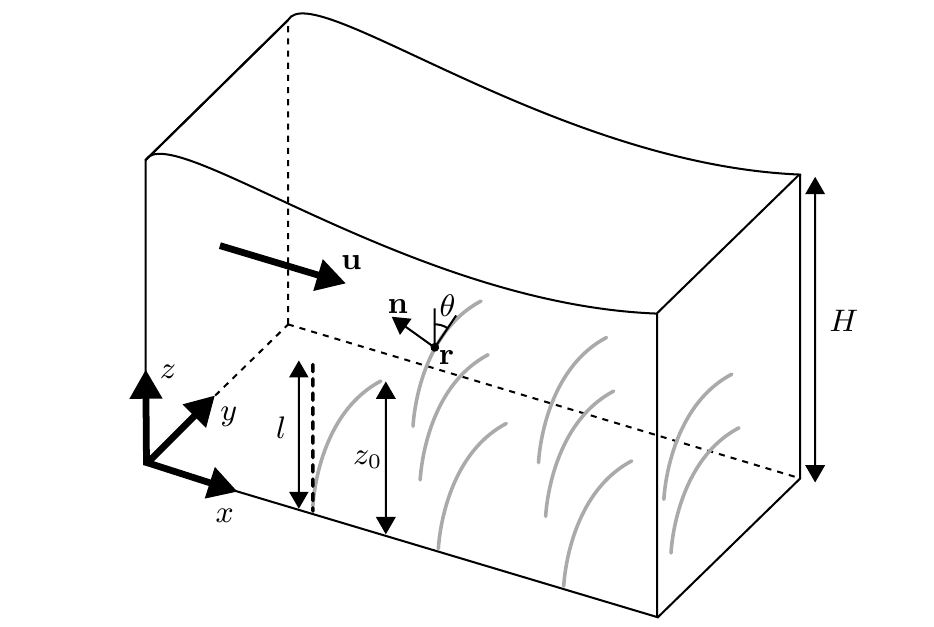}
	\end{overpic}}
	\caption{Schematic diagram of flow through a canopy. The fluid domain has depth $H$ and flow velocity, $\mathbf{u}$. The grey vegetative element with the centreline at $\mathbf{x}=(x,y,z)=\mathbf{r}$ denotes a single flexible plant that is rooted to the bed. The configuration is parametrised by arc length, $s$, and local angle of deflection, $\theta$. The vertical dashed line with length $l$ denotes the position of the element when it is load free and $z_0$ is the height of the plant in the deformed configuration.}
	\label{fgr:schematic}
\end{figure}

Consider a three-dimensional domain with fluid contained between $0 \leq z \leq H(x,y,t)$. The bed, at $z = 0$, is covered by a fully-submerged vegetative canopy that consists of $N$ identical plants of length $l$, which shall later be specified as linear elastic beams. A schematic of the setup and the coordinate system is given in \fig{schematic}. 

\subsection{Equations for the fluid}
\label{sec:equations_for_the_fluid}
\noindent In a more complete model, the location of each physical plant surface must be calculated as part of the problem. However, provided that the cross-sectional length scale of the plant is much smaller than the separation of neighbouring plants, we can consider the far-field approximation of the drag due to individual plants using a distribution of point forces in the Navier-Stokes equations. We thus consider an incompressible fluid in three dimensions with velocity $\mathbf{u}=(u,v,w)$ at time $t$ satisfying
\begin{align}
\nabla \cdot \mathbf{u} &= 0,\label{dim_ns1}\\
\rho\left(\pfrac{\mathbf{u}}{ t}+\mathbf{u}\cdot \nabla \mathbf{u}\right) &= -\nabla p -\rho 
g\hat{\mathbf{e}}_z + \mu \nabla^2\mathbf{u} + \mathbf{F},\label{dim_ns2}
\end{align}
where $\rho$ is the density of water, $p$ is pressure, $g$ is acceleration due to gravity and $\mu$ is the dynamic viscosity of water. The additional sink term, $\mathbf{F}$, that appears in the momentum equation \eqref{dim_ns2} incorporates the contribution of the $N$ plants; the precise form of $\mathbf{F}$ is discussed below. 

We model each of the $N$ plants as an inextensible linearly elastic beam with diameter $b$, which undergoes pure bending in the $xz$-plane. We parametrise the centreline of the $k$-th beam by $\mathbf{x}=\mathbf{r}^k(s,\tau)$ where $s$ is arc length and $\tau=t$ is time. We use $\tau$ rather than $t$ to emphassise the implicit change of variables from $(\mathbf{x},t)$ to $(k,s,\tau)$; this is a Lagrangian description of the canopy. We shall return to this later in \S\ref{ssec:coord} where we discuss the relationship between Eulerian and Lagrangian frames. 

We assume that the collective drag term can be written as
\begin{equation}\label{F}
\mathbf{F} \equiv \sum_{k=1}^N \mathbf{F}^k(\mathbf{x},t),
\end{equation}  
with each individual term, $\mathbf{F}^k$, accounting for the drag on the $k$-th plant. We suppose $\mathbf{F}^k$ depends on the difference between the fluid velocity and the beam velocity $\partial \mathbf{r}^k/\partial \tau$. In particular, we take
\begin{equation} \label{Fkind}
\mathbf{F}^k = -\frac{1}{2}\rho bC_D \int_0^{l}  \delta(\mathbf{x}-\mathbf{r}^k)\left(\mathbf{u}-\pfrac{ \mathbf{r}^k}{\tau}\right)\cdot\mathbf{n}^k \left|\left(\mathbf{u}-\pfrac{ \mathbf{r}^k}{\tau}\right)\cdot\mathbf{n}^k\right|\mathbf{n}^k\ \upd s.
\end{equation}
Here, $\delta$ denotes the Dirac delta function, $C_D$ is the drag coefficient and $\mathbf{n}^k$ is the upstream normal of the $k$-th beam's centreline in the $xz$-plane (cf. \fig{schematic}); if $\theta$ is the local angle of deflection at $s$, as measured from the upward vertical, then $\mathbf{n}^k=(-\cos\theta_k,0,\sin\theta_k)$. 

Our drag law \eqref{Fkind} can be interpreted as follows: for each element of arc length $\upd s$, the drag $\upd\mathbf{F}^k$ is that a tiled cylinder of length $\upd s$ would experience \citep{Sumer2006}. It has been proven both experimentally and numerically that this drag law is accurate until the beam approaches a configuration parallel to the flow \citep{Ramberg1983, Vakil2009,Zhao2009}. Analogous expressions for drag have been used by \citet{Luhar2016} for steady-state flow and by \citet{Leclercq2018} for unsteady flow. In the case of rigid and vertical beams ($\theta\equiv 0$), \eqref{Fkind} reduces to the formulation used in \citet{Singh2016}. We refer readers to \citet{Zhou2010} for a more extensive review of drag on tilted cylinders.

\subsection{Equations for the vegetation}
\label{sec:equations_for_the_vegetation}
\noindent Each individual plant is modelled as a linearly elastic beam with one end clamped perpendicularly to the substrate and the other end left free. For each beam, let $\mathbf{T}^k= (T_1^k, T_2^k, T_3^k)$ be the internal stress and $\mathbf{M}^k$ be the moment on a cross-section given by position vector $\mathbf{r}^k$. By considering the momentum balance on this cross-section which has infinitesimal thickness, $\mathbf{T}^k$ satisfies \citep{Landau1970}
\begin{align}
\pfrac{\mathbf{T}^k}{s} + \mathbf{q}^k &=\frac{\rho_B\upi b^2}{4} \pnfrac{2}{\mathbf{r}^k}{\tau},\label{dim_beam1}
\end{align}
where $\rho_B$ is the density per unit volume of the beam and $\mathbf{q}^k$ is the external force per unit length on the beam. 

We also consider the angular momentum balance on this cross-section. Noting that $\partial\mathbf{r}^k/{\partial s}=(\sin\theta_k,0,\cos\theta_k)$ is the local tangent vector of the plant, if we assume that the standard constitutive relation between $\mathbf{M}^k$ and $\mathbf{r}^k$ holds under dynamic conditions, namely \citep{Landau1970}
\begin{equation}
\mathbf{M}^k = EI \pfrac{\mathbf{r}^k}{s}\times \pnfrac{2}{\mathbf{r}^k}{s},
\end{equation}
then $\theta$ satisfies \citep{Howell2008}
\begin{align}
EI \pnfrac{2}{\theta_k}{s} + T_1^k \cos\theta_k -  T_3^k \sin \theta_k &= \rho_BI \pnfrac{2}{\theta_k}{\tau}.\label{dim_EI}
\end{align}
In this equation, $E$ is the Young's modulus and $I=\upi b^4/64$ is the moment of inertia of the cross-section about the $y$--axis. 

Finally, we must determine the load on the $k$-th plant, $\mathbf{q}^k$, in our fluid-structure interaction problem. By neglecting the effect of buoyancy in this analysis (by assuming that $\rho_B=\rho$), the total load is solely due to the drag in \eqref{Fkind}, so that
\begin{align}
\mathbf{q}^k &= \frac{1}{2} \rho bC_D \int_{\Omega} \delta(\mathbf{x}-\mathbf{r}^k)\left(\mathbf{u}-\pfrac{ \mathbf{r}^k}{\tau}\right)\cdot\mathbf{n}^k \left|\left(\mathbf{u}-\pfrac{ \mathbf{r}^k}{\tau}\right)\cdot\mathbf{n}^k\right|\mathbf{n}^k\ \upd\mathbf{x}
\end{align}
where $\Omega$ is the fluid domain. 
\subsection{Homogenisation of the canopy}
\label{sec:homo}
\noindent In real-world scenarios, the canopy is a medium with a complex micro-structure. Even with the simplifications we have made so far, it is impractical to monitor the positions and effects of individual plants in the flow. Instead, we use a simpler averaged model in which the canopy is an effective medium which contributes a bulk volumetric drag term (see \emph{e.g.} \citet{Nepf2012} and the references therein for previous models in this direction). Here, we briefly present the derivation of such a model by volume averaging. 

\begin{figure}
	\centering
	\begin{overpic}[width=0.7\textwidth]{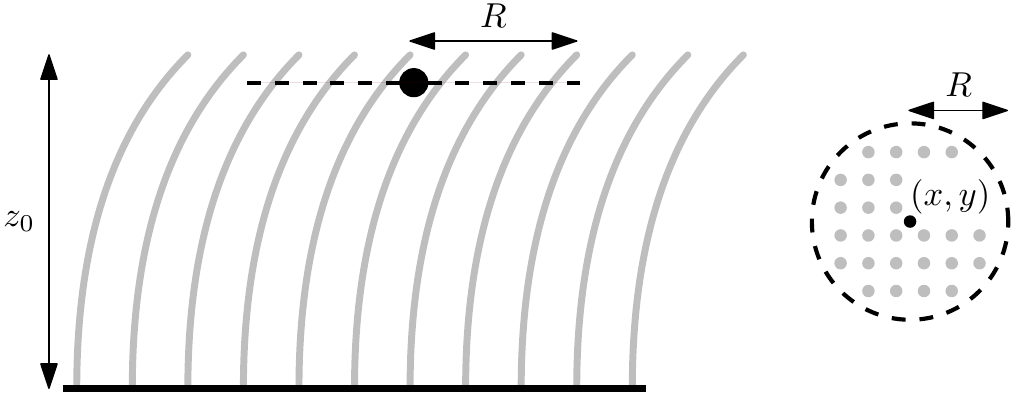}
		\put(0,36){(\textit{a})}
		\put(78,29.5){(\textit{b})}
	\end{overpic}
	\caption{Schematic diagram of a circle of radius $R$ (dashed), centred at $\mathbf{x}=(x,y,z)$, encircling a collection of plants (in grey) in the canopy. (\textit{a}) Side-view of the circle. (\textit{b}) Top-view of the circle.}
	\label{fgr:homo}
\end{figure} 

Consider a fixed $\mathbf{x} = (x, y, z)$. We define $\overline{\mathbf{F}}_R$ as the local average of the collective sink over a disk of radius $R$, namely
\begin{equation}
\overline{\mathbf{F}}_R(\mathbf{x},t)  = \frac{1}{\upi R^2}\iint_{C_R(x,y; z)} \mathbf{F}(\mathbf{x}',t)~\upd x'\upd y',
\end{equation}
where $C_R(x,y; z)$ is the two-dimensional disk of radius $R$ centred at the point $\mathbf{x}$ \emph{i.e.}
\begin{equation}
C_R(x,y; z) = \{(x',y',z): \, (x - x')^2 + (y - y')^2 \leq R\}
\end{equation}
(see \fig{homo}). By the definition of $\mathbf{F}$ in \eqref{F},
\begin{equation}\label{RHS}
\overline{\mathbf{F}}_R(\mathbf{x},t)  =  \frac{1}{\upi R^2}\iint_{C_R} \left[ \sum_{k=1}^N  \int_0^{l}Q_R^k(\mathbf{x}',t)\delta(\mathbf{x}'-\mathbf{r}^k)\mathbf{n}^k~\upd s\right]~\upd x' \upd y',
\end{equation}
where
\begin{equation}
Q_R^k(\mathbf{x}',t) = -\frac{1}{2}\rho bC_D \left(\mathbf{u}(\mathbf{x}',t)-\pfrac{ \mathbf{r}^k}{\tau}\right)\cdot\mathbf{n}^k \left|\left(\mathbf{u}(\mathbf{x}',t)-\pfrac{ \mathbf{r}^k}{\tau}\right)\cdot\mathbf{n}^k\right|.
\end{equation}
Now, by definition, 
\begin{equation}
\delta(\mathbf{x}' - \mathbf{r}^k) = \delta(x' - r_1^k)\delta(y' - r_2^k)\delta(z - r_3^k).
\end{equation}
In order to sample plants that pass through $C_R$, we rewrite the $s$--integral in \eqref{RHS} with respect to $z$. Assuming that the beam configurations do not overturn, 
\begin{equation}\label{s_z_CofV}
\frac{\upd}{\upd s} = \frac{\upd z}{\upd s}\frac{\upd }{\upd z}  = \cos\theta_k\frac{\upd }{\upd z}.
\end{equation}
Therefore,
\begin{align}
\overline{\mathbf{F}}_R(\mathbf{x},t)  
& = \frac{1}{\upi R^2}\sum_{k=1}^N Q_R^k(\mathbf{r}^k,t)\mathbbm{1}_k(\mathbf{x},t)\sec\theta_k(z,t)\mathbf{n}^k
\end{align}
where $\mathbbm{1}_k$ is an indicator function with $\mathbbm{1}_k(\mathbf{x},t)=1$ if the $k$-th plant passes through $C_R$ (and zero otherwise). 

We now consider the limit in which $R\rightarrow 0$ (with the neighbouring plant separation also tending to zero so that there are still many plants crossing $C_R$). Since we have assumed that the plants are identical in the canopy, we approximate $\mathbf{u}$ to be uniform in $C_R$ and similarly for the configuration and motion of the plants. With reference to the local averaging argument presented by \citet{Chapman1995}, we deduce the continuum approximation of $\mathbf{F}$ as
\begin{align}\label{F_homo}
\overline{\mathbf{F}}
= 
\lim_{R \to 0}\overline{\mathbf{F}}_R&=  -\frac{\rho bC_D\bar{N}}{2\cos\theta}\mathrm{H}(z_0-z)  \left(\mathbf{u}-\pfrac{ \mathbf{r}}{\tau}\right)\cdot\mathbf{n} \left|\left(\mathbf{u}-\pfrac{ \mathbf{r}}{\tau}\right)\cdot\mathbf{n}\right|\mathbf{n}
\end{align}
where $\bar{N}(x,y)$ is the number of plants planted per unit area (along the bed) and $\mathrm{H}$ is the Heaviside step function. The dependence of $\overline{\mathbf{F}}$ on $\sec\theta$ physically corresponds to the perpendicular distance between neighbouring plants being reduced when they tilt, so that the effective density of the canopy increases. For the remainder of this work, we will consider uniform canopies where $\overline{N}$ is constant. 


\subsection{Simplification and non-dimensionalisation} 
\label{ssec:2_non-dim}
\noindent Before continuing, we highlight two key simplifying assumptions that we will make. Following the approach by \citet{Dupont2010} and \citet{Singh2016}, we neglect the free surface that exists in reality and instead, treat the water depth $H$ to be constant and impose a shear-free condition along the top of the domain. Secondly, we impose a shear-free condition along the substrate by focusing on regimes with the Reynolds number based on a single element (plant) being much larger than unity \citep{Dunn1996,Ghisalberti2002}. We shall revisit some of these aspects in the final discussion in \S\ref{sec:8_discuss}.

We non-dimensionalise using the variables in our problem with the following scales:
\begin{equation}
[x]=[z]=[s]=H, \quad [\mathbf{u}]= U, \quad [t] = \frac{H}{U}, \quad [p] = \rho U^2, \quad [\mathbf{T}] = \rho bC_DH U^2.
\end{equation}
Foreseeing the calculations ahead in our stability analysis, we take $U$ to be the velocity at the top of the domain for steady unidirectional flows, which we will specify in \S\ref{sec:3_uni}. For the beam equations, we will neglect the inertial terms in the beam equations, \eqref{dim_beam1} and \eqref{dim_EI}, by focusing on the regimes in which $\pi b/(4C_D H)\ll 1$ and $\rho U^2/E\ll 1$. The former inequality is a good approximation for slender vegetation with $b/H \ll 1$. The latter inequality holds whenever the velocity of the fluid is small compared to the speed of sound through the beams, which is also typical for aquatic vegetation \citep{Lei2016}. 

With all the variables being henceforth dimensionless, we
have following system of dimensionless equations:
 \begin{subequations} \label{eq:mainsys}
 \begin{eqnarray}
 \nabla\cdot \mathbf{u} &=& 0, \label{eq1}\\
 \hspace{-0.3cm}\frac{\partial \mathbf{u}}{\partial t} + \mathbf{u}\cdot\nabla \mathbf{u} &=&
  -\nabla p + \dfrac{1}{\Rey}\nabla^2\mathbf{u}-\dfrac{1}{Fr^2}\hat{\mathbf{e}}_z \nonumber\\
 &&-\frac{\lambda\mathrm{H}(z_0-z)}{2\cos\theta}\left(\mathbf{u}-\pfrac{ \mathbf{r}}{\tau}\right)\cdot\mathbf{n} \left|\left(\mathbf{u}-\pfrac{ \mathbf{r}}{\tau}\right)\cdot\mathbf{n}\right|\mathbf{n},
 \label{eq2}\\ 
 \pfrac{\mathbf{T}}{s} &=& - \frac{1}{2}\left(\mathbf{u}-\pfrac{ \mathbf{r}}{\tau}\right)\cdot\mathbf{n} \left|\left(\mathbf{u}-\pfrac{ \mathbf{r}}{\tau}\right)\cdot\mathbf{n}\right|\mathbf{n},\label{eq3}\\
 \pnfrac{2}{\theta}{s} &=& C_Y(- T_1 \cos\theta + T_3 \sin \theta),
 \label{eq4}
 \end{eqnarray}
 with the boundary conditions 
 \begin{align}	
\fbox{\small no penetration}&\quad w\rvert_{z=0,1}=0,\label{bc1}\\  
\fbox{\small no shear}&\quad\left[\pfrac{u}{z}+\pfrac{w}{x}\right]_{z=0,1}=0,\  \left[\pfrac{v}{z}+\pfrac{w}{y}\right]_{z=0,1}=0,\label{bc2}\\
\fbox{\small cantilever beam}&\quad \mathbf{T}\rvert_{s=h}=\mathbf{0},\
\theta\rvert_{s=0}=0,\  \pfrac{\theta}{s}\biggr\rvert_{s=h}=0,\  \label{bc3}\\
\fbox{\small constraint on $z_0$}& \quad  \int_0^{z_0} \sec\theta~\upd{z} =h,\label{bc4}
 \end{align}
 \end{subequations}
 
\begin{table}
	\begin{center}
		\def~{\hphantom{0}}
		\begin{tabular}{rcc}
			& Symbol      & Expression    \\[2pt]
		Reynolds number         & $\Rey$      & $\rho UH/\mu$             \\[2pt]
		Froude number           & $Fr$          & $U/\sqrt{gH}$ \\[2pt]
		Submergence ratio       & $h$ & $l/H$                     \\[2pt]
		Canopy density          & $\lambda$   & $C_D\bar{N}bH$                    \\[2pt]
		Cauchy number           & $C_Y$       & $\rho bC_D H^3U^2/(EI)$\\    
		\end{tabular}
		\caption{A summary of the dimensionless parameters in the governing equations of flow through a homogenised canopy \eqref{eq:mainsys}.}
		\label{table:dimless_param}
	\end{center}
\end{table}
where the dimensionless parameters $\Rey$, $Fr$, $h$, $\lambda$, and $C_Y$ are defined in table~\ref{table:dimless_param}. In particular, $\lambda$ is the product of the dimensionless planting density, $\bar{N}H^2$, the aspect ratio, $b/H$,  and the geometry factor, $C_D$. Finally, $C_Y$ characterises the balance between static deflection and loading due to drag. We have chosen the convention where we incorporate the geometry of the plant by scaling $C_Y$ with $C_D$ and $I$ instead of introducing a slenderness parameter \citep{DeLangre2008}. 

We have seen in \eqref{eq:mainsys} that it is natural to write the Navier-Stokes equations in Eulerian coordinates but it is more natural to write the beam equations in body-fitted coordinates (\emph{i.e.} arc length $s$). These systems are related via $\partial \mathbf{r}/\partial s = (\sin\theta,0,\cos\theta)$. We examine the translation between the two coordinate systems more fully in \S\ref{ssec:coord}.

We presented a complicated system of coupled non-linear partial differential equations for the model of the fluid-vegetation system. The purpose of this work is to analyse the instability mechanism of monami. Therefore, in the next section, we will develop tractable unidirectional reductions of the governing equations so that steady configurations are readily solvable. 

%
%
%

\section{Steady unidirectional flow}
\label{sec:3_uni}
In this section, we seek solutions of the governing system \eqref{eq:mainsys} where the flow is steady and unidirectional along the $x$--axis (the streamwise direction), with 
\begin{equation}
\mathbf{u}=u(z)\hat{\mathbf{e}}_x.
\end{equation}
The velocity only depends on the distance from the substrate. Historically, field studies, controlled experiments, and more recently, numerical simulations typically measure $u(z)$ [cf. \citet{Nepf2012}]. To account for eddies in the flow, we follow the approach by \citet{Singh2016} and assign a constant eddy viscosity, $\nu^*$, to replace $\mu/\rho$ in \eqref{dim_ns2}. Thus, for the remainder of this work, we use the effective Reynolds number,
\begin{equation}
\Rey^* = \frac{UH}{\nu^*},	
\label{eddy}
\end{equation}
to characterise unidirectional flows. In particular, the characteristic velocity, $U$, is chosen such that $u(1)=1$ (see discussion in \S\ref{ssec:2_non-dim}). 


\subsection{Theoretical formulation}
\label{sec:theoretical_formulation}
\noindent Since the flow is unidirectional, we can deduce from the momentum equation of the fluid \eqref{dim_ns2} that the flow is driven by a constant pressure gradient, $\px>0$. The governing equations \eqref{eq:mainsys} reduce to the following system of differential equations:
\begin{subequations}\label{eq:steady}

\begin{align}
\frac{\upd^2 u}{\upd z^2} &=
-\Rey^*\px+\mathrm{H}(z_0-z)\frac{\Rey^*\lambda}{2}u\cos\theta|u\cos\theta|, \label{z_u}\\
\frac{\upd T_1}{\upd s} &= -\dfrac{1}{2}u\cos^2\theta|u\cos\theta|,\\
\frac{\upd T_3}{\upd s} &= \dfrac{1}{2}u\sin\theta\cos\theta|u\cos\theta|,\\
\frac{\upd^2 \theta}{\upd s^2} &= C_Y(-T_1\cos\theta+T_3\sin\theta),
\label{z_theta} 
\end{align} 
\end{subequations}
with $\px$ chosen such that $u(1)=1$. To solve for the steady configuration, we rewrite the equations \eqref{eq:steady} as a system of ordinary differential equations in $z$ and solve the equations numerically [with the corresponding boundary conditions in \eqref{eq:mainsys}] in MATLAB using the boundary value problem solver \texttt{bvp4c}. Once we obtain the solution, the centreline of the homogenised plant configuration, $\mathbf{r}$, can be determined in Cartesian coordinates from the relation
\begin{equation}
\Dfrac{x}{z}=\tan\theta.
\label{dxdz}
\end{equation}
As an aside, we note that our model predicts a parabolic flow profile above the canopy rather than logarithmic as in classic boundary layer flows \citep{Nikora2010}. Since the canopy enhances flow mixing above the canopy, it has been experimentally
shown that the logarithmic scaling is only recovered when $z \gtrapprox 2h$ [see \citet{Sharma2018} and references therein]. Provided that $h=O(1)$, which is typical for aquatic vegetation \citep{Nepf2012}, we can assume that we are not in the logarithmic regime. For flows with $h\ll1$, the transition zone where $h\leq z \lessapprox 2h$ is known as the roughness sublayer \citep{Finnigan2000}.

\subsection{Unidirectional flows through rigid canopies ($C_Y = 0$)} 
\label{ssec:rigid_canopy}
\noindent In the case of flow through a rigid canopy, the momentum equation of the fluid naturally decouples from the beam equations at leading-order. Below, we present the asymptotic analysis in the limits of sparse and dense canopies. 

\subsubsection{Asymptotic approximation of $u$ for sparse canopies ($\lambda \ll 1$)}
\label{subsec:sparse_canopy}
\begin{figure}
	\centering
	\includegraphics[width=\textwidth]{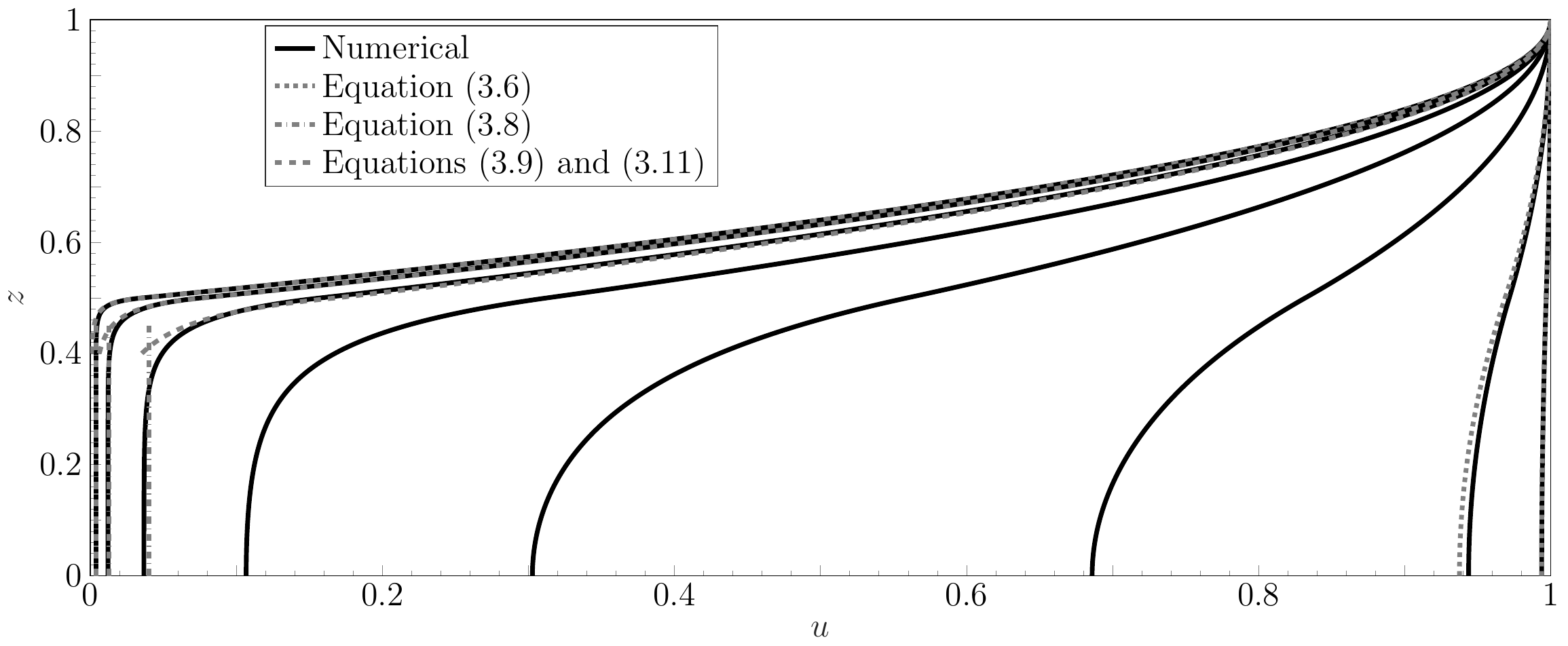}
	\caption{The effect of increasing canopy density to steady unidirectional flows through a rigid canopy when $\Rey^*=10^3$ and $h=0.5$. Flow profiles are shown in black solid lines with grey lines indicating asymptotic approximations as described in \S\ref{ssec:rigid_canopy}.}
	\label{fgr:steady_lambda}
\end{figure}

To begin, observe the velocity profiles of \fig{steady_lambda}, shown at different values of the canopy density parameter. In the limit the canopy density tends to zero, the velocity profile approaches that corresponding to a uniform and unobstructed flow. Noting that a more convenient perturbation parameter is $\Rey^*\lambda$, we let $u=1+\Rey^*\lambda\tilde{u}$ and $\px=\lambda\tilde{\px}$. Then by \eqref{z_u}, we have
\begin{equation}
\Dnfrac{2}{\tilde{u}}{z} \sim 
\begin{cases}
-\tilde{\px}+\dfrac{1}{2},&\mathrm{if}~z\leq h,\\
-\tilde{\px},&\mathrm{if}~z>h.
\end{cases}
\end{equation} 
By solving for $\tilde{u}$ (and $\tilde{P}$), we deduce that 
\begin{equation}
u=1+\Rey^*\lambda\tilde{u}\sim\begin{cases}
1-\dfrac{\Rey^*\lambda}{4}(1-h)(h-z^2),&\mathrm{if}~z\leq h,\\
1-\dfrac{\Rey^*\lambda }{4}h\left(z-1\right)^2,&\mathrm{if}~z>h,
\end{cases}
\label{asy_sparse}
\end{equation}
and this confirms the behaviour illustrated in \fig{steady_lambda}. In particular, we remark that as the canopy density increases, the flow velocity reduces everywhere due to drag; however, as expected, the reduction is greatest within the canopy itself. 



\subsubsection{Asymptotic approximation of $u$ for dense canopies $(\lambda \gg 1$)}
\label{subsec:dense_canopy}
%

%

We also observe from figure~\ref{fgr:steady_lambda} that in the dense-canopy limit, $\lambda \gg 1$, the velocity, $u(z)$, is apparently divided into two outer regions ($z < h$ and $z > h$), as well as an inner region near $z = h$. In the limit $\lambda\rightarrow \infty$, we observe that the flow becomes approximately uniform in the canopy, where $0 \leq z \leq h$ and $h - z = O(1)$. From \eqref{z_u}, we have that in this region, 
\begin{equation}
 \frac{\upd^2u}{\upd \hat{z}^2} = -\Rey^*\px + \frac{\Rey^*\lambda}{2}u|u|,
\label{z_u_asy2}
\end{equation}
and thus to leading order, the pressure gradient balances the drag, and the velocity below the canopy satisfies
\begin{equation}\label{asy_dense_outer}
u_\text{below}(z) \sim \left(\frac{2\px}{\lambda}\right)^{1/2} \quad \text{for $0 \leq z \leq h$ and $h-z = O(1)$}.
\end{equation}
This matches the result of \citet[\S{5}]{Poggi2004a} and \citet[\S{3}]{Singh2016}. Before deriving the solution in the boundary layer, we note that the solution for $z\geq h$  can be found exactly: integrating \eqref{z_u} and applying the surface boundary conditions, $u(1)=1$ and $u'(1)=0$, we find that for the solution above the canopy, 
\begin{equation}
\label{asy_dense_above} 
u_\text{above}(z) = \Rey^*\px\left(z-\frac{z^2}{2}-\frac{1}{2}\right) + 1 \quad \text{for $z \geq h$.}
\end{equation}
Note that the constant quantity, $\px$, itself must be expanded in $\lambda$ and will be determined through boundary conditions. In the inner region, we substitute $z = h-\lambda^{-1/3}\eta$ and $u = \lambda^{-1/3}\tilde{u}$ into \eqref{z_u} with $\eta\geq 0$. The inner solution satisfies 
\begin{equation} \label{asy_dense_eqn}
\left(\Dfrac{\tilde{u}}{\eta}\right)^{2} \sim \frac{\Rey^*}{3}\tilde{u}^3 + C,
\end{equation}
where the integration constant $C = 0$ in order to match \eqref{asy_dense_outer} to this order of approximation. As a result, taking the relevant branch of \eqref{asy_dense_eqn} and requiring that the solution is continuous with \eqref{asy_dense_above}, we have that for the solution within the boundary layer,
\begin{equation} \label{asy_dense_inner}
u_\text{layer}(z) \sim  \left[\left(\frac{\Rey^*\lambda}{12}
\right)^{1/2}(h-z)+\left\{1-\frac{\Rey^*\px}{2}\left(1-h\right)^2\right\}^{-1/2}\right]^{-2},
\end{equation}
which is valid for $z \leq h$ and $h - z = O(\lambda^{-1/3})$. Finally, in order to determine the leading-order behaviour of the constant pressure gradient, $\px$, we require that the gradient of \eqref{asy_dense_above} matches that given by the boundary layer of \eqref{asy_dense_eqn} at $z = h$. This yields
\begin{equation} \label{eq:pxseries}
	\px  \sim \frac{2}{\Rey^*(1-h)^2 } \left(1- \left[\frac{12}{\Rey^*\lambda(1-h)^2}\right]^{1/3}\right).
\end{equation}
Note that in the above expression, we have retained the first two orders in $\px$ so as to ensure higher accuracy in the above-canopy solution for a wider range of $\lambda$ values. In \fig{steady_lambda}, we observe good agreement between the exact numerical flow profiles and our matched-asymptotic approximations given in \eqref{asy_dense_outer}, \eqref{asy_dense_above}, and \eqref{asy_dense_inner}.

\subsection{Unidirectional flows through flexible canopies ($C_Y \neq 0$)}
\label{ssec:flex_canopy}
\begin{figure}
	\centering
	\includegraphics[width=\textwidth]{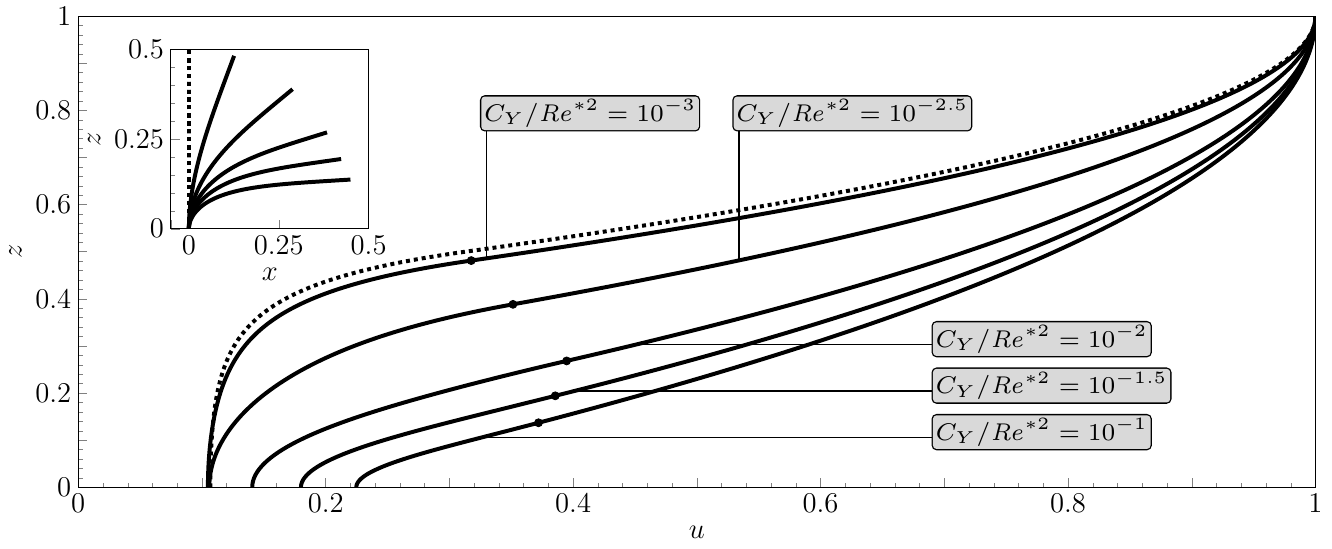}
	\caption{The effect of increasing the flexibility of vegetation to steady unidirectional flows when $\Rey^*=10^3$, $\lambda=1$ and $h=0.5$. Flow profiles are shown in solid lines for different values of $C_Y/\Rey^{*2}$, with the corresponding homogenised beam configuration (to scale) in the inset. Dots in the main figure indicate the respective height of the canopy and dotted lines indicate the configuration in the rigid canopy limit $C_Y/\Rey^{*2} = 0$.}
	\label{fgr:steady_flex}
\end{figure}

\noindent Having gained some intuition on flows through rigid canopies, we now explore the differences in flows through flexible canopies---the main motivation of this work. Before we continue, we first define how we will vary the flexibility. We recall from \S\ref{sec:2_model} that the Cauchy number, $C_Y$, characterises the amount of deflection due to drag. Hence, it varies with the velocity scale (cf. table~\ref{table:dimless_param}). In order to vary flexibility independently, we vary the ratio
\begin{equation}
\frac{C_Y}{\Rey^{*2}} = \frac{\rho C_DbH\nu^{*2}}{EI}
\end{equation}
in our analysis for the remainder of this work. Flow profiles for varying $C_Y/\Rey^{*2}$ and fixed canopy density are shown in \fig{steady_flex}.

As for flows through rigid canopies, every flow profile in \fig{steady_flex} increases monotonically in $z$ and inflects at the top of the canopy. As we increase the flexibility of the vegetation (by increasing $C_Y/\Rey^{*2}$), less of the domain is obstructed and we get faster flows at any given $z$. For applications such as flood control, if we use the (dimensional) maximum velocity as a simple measure for damage, the results suggest that upright obstacles attenuate a steady flow most effectively. We will revisit this conclusion in the next section when we consider the unsteady problem. 

As an aside, we note that for regimes where the canopy is dense ($\lambda\gg 1$), we expect the approximations in \S\ref{subsec:dense_canopy} to hold for plants that are sufficiently stiff. In this limit, since $u\sim \lambda^{-1/3}$, the load and hence the deflection of each plant is negligible. 

\section{Stability analysis of the steady configuration}
\label{sec:4_LSA}
Our main task in this section is to extend the stability analysis by \citet{Singh2016} on rigid canopies to flexible canopies. By modelling the canopy as an array of elastic beams, we are interested in the temporal evolution of both the perturbed flow and the canopy configuration. 

By considering the steady configurations of \S\ref{sec:3_uni} as base states, we first derive the system of equations the perturbations satisfy at leading order. We then impose a spectral decomposition on the perturbations and solve the corresponding generalised eigenvalue problem numerically. In this work, we consider two-dimensional disturbances in the $xz$--plane, which is sufficient if we are primarily interested in the critical conditions for instability \citep{Drazin2004}. 

\subsection{Relating Eulerian and Lagrangian frame}
\label{ssec:coord}
\noindent Before we consider perturbations of the base flow, we again note that the spatial variables of the fluid are in the Eulerian frame but the variables of the homogenised plant are parametrised by the arc length, $s$. Since it is natural for us to perform stability analysis in Cartesian coordinates, we first express the variables in the system \eqref{eq:mainsys} in terms of $(\mathbf{x},t)$. 

Firstly, although the transformation between $s$ and $z$ \eqref{s_z_CofV} still holds in the dynamic problem, namely
\begin{align}
s=\int_0^z \sec\theta~\upd \overline{z},
\label{s}
\end{align}
this transformation is perturbed when $\theta$ is perturbed and solved as part of the problem. Expanding all variables as $f=f^*+\ep \hat{f}+\dots$, where $f^*$ is the steady state and $\ep \hat{f}$ is the perturbation, and linearising in $\ep$, we find that  
\begin{align}
\quad s^*=\int_0^z \sec\theta^*~\upd \overline{z} 
\quad \text{and} \quad
\hat{s}=\int_0^z \sec\theta^*\tan\theta^*\hat{\theta}~\upd \overline{z}.
\refstepcounter{equation}\tag{\theequation $a$,\,$b$}
\label{s0s1}
\end{align}

Secondly, the Lagrangian time derivative, $\partial/\partial \tau$, is distinct from the Eulerian time derivative, $\partial/\partial t$. As a result, by \eqref{s},
\begin{align}
\pfrac{}{t} &= \pfrac{\tau}{t}\pfrac{}{\tau} + \pfrac{s}{t}\pfrac{}{s} = \pfrac{}{\tau}+\left(\pfrac{}{t}\int_0^z\sec\theta~\upd\overline{z}\right)\left(\cos\theta\pfrac{}{z}\right)\label{t}.
\end{align}
With the transformations above, we can now rewrite every dependent variable with respect to $(\mathbf{x},t)$. Before we write down the linearised equations, we need to determine the expansions of two key quantities: the velocity of the homogenised plant, $\partial \mathbf{r}/\partial \tau$, and the height of the canopy, $z_0$. 
\subsubsection{Plant velocity}
We would like to express the plant velocity $\partial\mathbf{r}/\partial\tau=(\partial r_1/\partial\tau,0,\partial r_3/\partial \tau)$,
at a fixed point on the plant in terms of $(\mathbf{x},t)$. In the Lagrangian frame, for a given $s$, 
\begin{equation}
\pfrac{r_1}{\tau}=\pfrac{}{\tau}\int_0^s \sin\theta~\upd\overline{s}=\int_0^s \cos\theta\pfrac{\theta}{\tau}~\upd\overline{s} = \int_0^z \pfrac{\theta}{\tau}~\upd\overline{z}. 
\end{equation}
It remains to express $\partial \theta/\partial \tau$ as a function of $(\mathbf{x},t)$. By \eqref{t}, 
\begin{equation}
\pfrac{\theta}{\tau} = \pfrac{\theta}{t}-\cos\theta\pfrac{\theta}{z}\pfrac{s}{t}.
\end{equation}
Therefore, 
\begin{equation}
\pfrac{r_1}{\tau}=\int_0^z  \pfrac{\theta}{t}-\cos\theta\pfrac{\theta}{z}\pfrac{s}{t}~\upd\overline{z}.
\end{equation}
Define $V_1=\partial r_1/\partial\tau$ and $V_3 = -\partial r_3/\partial\tau$. We have that $V_{1,3}^*=0$ for the base state, with 
\begin{equation}
\hat{V}_1= \int_0^z  \pfrac{\hat{\theta}}{t}-\cos\theta^*\Dfrac{\theta^*}{z}\pfrac{\hat{s}}{t}~\upd\overline{z}
\label{v1express}
\end{equation} 
and 
\begin{equation}
\hat{V}_{3}= \int_0^z  \tan\theta^*\left[\pfrac{\hat{\theta}}{t}-\cos\theta^*\Dfrac{\theta^*}{z}\pfrac{\hat{s}}{t}\right]~\upd\overline{z}.
\label{v3express}
\end{equation} 

\subsubsection{Height of the canopy}

Recall that the height of the canopy, $z_0$, also varies as we perturb the base state. This has to be taken into account when we impose boundary conditions on the differential equations of the perturbations at the original height. 

Using the expansion of $\theta$ in the integral constraint on $z_0$ \eqref{bc4}, gives
\begin{equation}\label{z_0}
\hat{z}_{0}=-\cos\theta^*(z_0^*)\int_0^{z_0^*} \sec\theta^*\tan\theta^*\hat{\theta}~\upd z.
\end{equation}
Having derived for expressions of $\hat{V}_{1}$, $\hat{V}_{3}$ and $\hat{z}_{0}$, we now proceed to derive the system of equations the perturbations satisfy. 

\subsection{Linear stability analysis}
\noindent By substituting in the perturbed base state into the original system \eqref{eq:mainsys} and collecting the linear terms, we find that the perturbations satisfy 
\begin{subequations}\label{eq:first_order}
\begin{eqnarray}
\pfrac{\hat{u}}{x}+\pfrac{\hat{w}}{z} &=& 0,\label{lsa_first1}\\
\pfrac{\hat{u}}{t}+u^*\pfrac{\hat{u}}{x}+\hat{w}\Dfrac{u^*}{z} &=&-\pfrac{\hat{p}}{x} + \frac{1}{\Rey^*}\nabla^2\hat{u}-\lambda \mathrm{H}(z_0^*-z)u^*\cos\theta^*\hat{\alpha} \nonumber\\
&&-\frac{\lambda}{2}\delta(z-z_0^*)u^{*2}\cos^2\theta^*\hat{z}_{0},\label{lsa_first2}\\
\pfrac{\hat{w}}{t}+u^*\pfrac{\hat{w}}{x} &=&
-\pfrac{\hat{p}}{z} + \frac{1}{\Rey^*}\nabla^2\hat{w} +\lambda \mathrm{H}(z_0^*-z)\left( \frac{1}{2}u^{*2}\hat{\theta} +  u^*\sin\theta^*\hat{\alpha}\right)\nonumber\\
&&+\frac{\lambda}{2}\delta(z-z_0^*)u^{*2}\sin\theta^*\cos\theta^*\hat{z}_{0},\label{lsa_first3}\\
\pfrac{\hat{T}_{1}}{z} &=&- u^*\cos\theta^*\hat{\alpha},\label{lsa_first4}\\
\pfrac{\hat{T}_{3}}{z} &=&\frac{1}{2}u^{*2}\hat{\theta}+ u^*\sin\theta^*\hat{\alpha},\label{lsa_first4a}\\
\pnfrac{2}{\hat{\theta}}{ z}  - 2\tan\theta^*\Dfrac{\theta^*}{z}\pfrac{\hat{\theta}}{z}  
&=&\sec^2\theta^*\left(\Dfrac{\theta^*}{z}\right)^2\hat{\theta}-C_Y\sec\theta^*\left(\hat{T}_{1}+T_1^*\tan\theta^*\hat{\theta}\right)\nonumber\\
&&+C_Y\sec\theta^*\left[\tan\theta^*\hat{T}_{3}+T_{3}^*(2\sec^2\theta^*-1)\hat{\theta}\right]\label{lsa_first5}
\end{eqnarray}
where
\begin{equation}
\hat{\alpha} = \cos\theta^*\left(\hat{u}-\hat{V}_{1}\right)-\sin\theta^*\left(u^*\hat{\theta}+\hat{w}+\hat{V}_{3}\right).
\end{equation}
\end{subequations}
We will give the boundary conditions explicitly in the next section when we consider travelling wave perturbations. 

\subsection{Travelling wave perturbations}
\label{ssec:travel}
\noindent Anticipating the calculations ahead, it is convenient to first define the stream function of the flow, $\psi=\psi^*+\ep\hat{\psi}+\dots$, such that
\begin{equation}
u^* = \Dfrac{\psi^*}{z} \qquad\mathrm{and}\qquad 
\begin{pmatrix} \hat{u}\\ \hat{w}\end{pmatrix} =\begin{pmatrix} \partial\hat{\psi}/\partial z\\ - \partial\hat{\psi}/\partial x\end{pmatrix}.
\end{equation}
By rewriting $\hat{u}$ and $\hat{w}$ in \eqref{lsa_first1}--\eqref{lsa_first3} as derivatives of $\hat{\psi}$, the incompressibility condition \eqref{lsa_first1} is then naturally satisfied. Furthermore, by equating the derivatives of $\hat{p}$, we can combine \eqref{lsa_first2}--\eqref{lsa_first3} into a single differential equation for $\hat{\psi}$. With $\hat{u}$, $\hat{w}$, and $\hat{p}$ being eliminated, we consider a Fourier decomposition in the streamwise direction by letting
\begin{equation}\label{lsa_wave} (\hat{\psi},\hat{\theta},\hat{T}_{1},\hat{T}_{3})=(\phi(z),\vartheta(z),\mathcal{T}_1(z),\mathcal{T}_3(z))\mathrm{e}^{\mathrm{i}kx+\sigma t},
\end{equation}
real part understood, with $k$ being the wavenumber of the perturbation along the domain and $\sigma$ being the eigenvalue of the problem. By substituting the ansatz \eqref{lsa_wave} into \eqref{eq:first_order} and use primes ($'$) to denote derivatives in $z$, we find
\begin{subequations}\label{lsa_final}
\begin{align}
\frac{\phi''''-2k^2\phi''+k^4\phi}{\Rey^*} -(\sigma+\mathrm{i}ku^*)(\phi''-k^2\phi)+\mathrm{i}k{u^{*}}''\phi&= \begin{cases}\mathcal{S}_\mathrm{c}'+\mathrm{i}k\mathcal{S}_\mathrm{s},&\mathrm{if}~z\leq z_0^*,\\0, &\mathrm{if}~z>z_0^*,
\end{cases}
\label{lsa_final1}
\end{align}
\begin{eqnarray}
\mathcal{T}_{1}' &=&-u^*\cos\theta^*\left[\cos\theta^*\left(\phi'-\sigma\mathcal{A}\right)-\sin\theta^*\left(u^*\vartheta-\mathrm{i}k\phi+\sigma\mathcal{B}\right)\right],\label{lsa_final2}\\
\mathcal{T}_{3}' &=&\frac{1}{2}u^{*2}\vartheta+u^*\sin\theta^*\left[\cos\theta^*\left(\phi'-\sigma\mathcal{A}\right)-\sin\theta^*\left(u^*\vartheta-\mathrm{i}k\phi+\sigma\mathcal{B}\right)\right],\label{lsa_final3}\\
\vartheta''
&=&2\tan\theta^*{\theta^*}'\vartheta'  + \sec^2\theta^*\left({\theta^*}'\right)^2\vartheta-C_Y\sec\theta^*\left(\mathcal{T}_1+T_{1}^*\tan\theta^*\vartheta\right)\nonumber\\
&&+C_Y\sec\theta^*\left[\tan\theta^*\mathcal{T}_3+T_{3}^*(2\sec^2\theta^*-1)\vartheta\right],
\label{lsa_final4}
\end{eqnarray}

where
\begin{align}
\mathcal{C}&=\int_0^{z}\sec\theta^*\tan\theta^*\vartheta~\upd\overline{z},\label{lsa_C}\\\mathcal{A} &= \int_0^z  \vartheta-\mathcal{C}\cos\theta^*{\theta^*}'~\upd\overline{z},\label{lsa_A}\\
\mathcal{B} &= \int_0^z \tan\theta^*\vartheta-\mathcal{C}\sin\theta^*{\theta^*}'~\upd\overline{z},\label{lsa_B}\\
\mathcal{S}_\mathrm{c} &= \lambda u^*\cos\theta^*\left[\cos\theta^*\left(\phi'-\sigma\mathcal{A}\right)-\sin\theta^*\left(u^*\vartheta-\mathrm{i}k\phi+\sigma\mathcal{B}\right)\right],\\
\hspace{-2cm} \mathcal{S}_\mathrm{s}& = \lambda\left[\frac{1}{2}u^{*2}\vartheta+u^*\sin\theta^*\left[\cos\theta^*\left(\phi'-\sigma\mathcal{A}\right)-\sin\theta^*\left(u^*\vartheta-\mathrm{i}k\phi+\sigma\mathcal{B}\right)\right]\right],
\end{align}
are functions of $z$, $\theta$, and $\phi$. The stream function satisfies a modified Orr-Sommerfeld equation \citep{Drazin2004}, modified by additional momentum sinks, $\mathcal{S}_\mathrm{c}'$ and $\mathcal{S}_\mathrm{s}$, in \eqref{lsa_final1} that are only active in the obstructed part of the domain (at steady state). The expressions for these sinks are coupled to the perturbed beam equations to account for canopy deformation. The corresponding boundary conditions are 

\begin{align}
\fbox{\small fluid} &\qquad \phi(0)=0,~\phi(1)=0,~\phi''(0)=0,~\phi''(1)=0,\label{lsa_bc1}\\
\hspace{-0.5cm}\fbox{\small plant} &\qquad \mathcal{T}_1(z_0^*)=\left[-\frac{u^{*2}}{2}\mathcal{C}\cos^3\theta^*\right]_{z=z_0^*},~\mathcal{T}_3(z_0^*)=\left[\frac{u^{*2}}{2}\mathcal{C}\sin\theta^*\cos^2\theta^*\right]_{z=z_0^*},\\
&\qquad\vartheta(0)=0,~\vartheta'(z_0^*)=0. 
\label{lsa_bc2}
\end{align} 
Finally, although $\phi$ and $\phi'$ are continuous at $z=z_0^*$, due to the discontinuous momentum sink in \eqref{eq2}, we have

\begin{eqnarray}
\left[\phi''(z)\right]_{z_0^{*-}}^{z_0^{*+}} 
&=&
-\frac{\Rey^*\lambda}{2}\left[u^{*2}\mathcal{C}\cos^3\theta^*\right]_{z=z_0^*},
\label{lsa_phi''}\\
\left[\phi'''(z)\right]_{z_0^{*-}}^{z_0^{*+}}&=
& -\Rey^*\lambda \left[u^*\cos\theta^*\left\{\cos\theta^*\left(\phi'-\sigma\mathcal{A}\right)-\sin\theta^*\left(u^*\vartheta-\mathrm{i}k\phi+\sigma\mathcal{B}\right)\right\}\right]_{z=z_0^*}\nonumber\\ &&-\frac{\mathrm{i}k\Rey^*\lambda}{2}\left[u^{*2}\mathcal{C}\sin\theta^*\cos^2\theta^*\right]_{z=z_0^*}.
\label{lsa_phi'''}
\end{eqnarray}
\end{subequations}
Further discussion on how the jump conditions are derived can be found in Appendix~\ref{app:A}. For a given base state and a given $k$, we seek for values of $\sigma(k)$ such that there are non-trivial eigenmodes of the system of integro-differential equations \eqref{lsa_final} for $\phi$, $\vartheta$, $\mathcal{T}_1$, and $\mathcal{T}_3$. In particular, we are interested in the most unstable (or the least stable) mode in each spectrum. We solve this eigenvalue problem numerically. 

\section{Numerical results}
\label{sec:5_results}
In this section, we visualise some typical solutions for the unstable modes and compare our numerically determined eigenvalues to those of simplified problems with the hope of gaining some physical insights. We present a more extensive exploration of the solution space and discuss the critical parameters for the onset of stability in \S\ref{sec:6_crit}.

\subsection{Numerical method for solving the eigenvalue problem}
\label{ssec:LSA_num}
%
\noindent We first rewrite the eigenvalue problem \eqref{lsa_final} as a system of ordinary differential equations in $z$. In this setting, $\phi$ is defined in $[0,1]$ while other variables are only defined in $[0,z_0^*]$. Furthermore, $\phi$ has discontinuous derivatives at $z=z_0^*$. Therefore, we partition the domain into $[0,z_0^*]$ and $[z_0^*,1]$ and split $\phi$ into two separate functions, namely
\begin{equation}
\phi_{\mathrm{bot}}=\phi(z\leq z_0^*)\qquad\mathrm{and}\qquad \phi_{\mathrm{top}} = \phi(z\geq z_0^*).
\end{equation}
Following the practice of previous work in solving Orr-Sommerfeld problems, we solve this system of equations numerically using a spectral method with Chebyshev polynomials of the second kind \citep{Orszag1971}. To implement this method, in each interval, we discretise the dependent variables in \eqref{lsa_final} into their function values at $N$ Chebyshev nodes. Note that $z=z_0^*$ is an edge node in each interval, which represents $z_0^{*-}$ and $z_0^{*+}$ when we impose the jump conditions \eqref{lsa_phi''}--\eqref{lsa_phi'''}. Once we have defined the nodes, we then construct the discrete version of the differential operators in \eqref{lsa_final} with \textit{Chebfun}, an open-source package \citep{Driscoll2014}. Finally, we can rearrange the discrete equations so that $\sigma$ satisfies a generalised eigenvalue problem of the form 
\begin{equation}
\mathsfbi{A}(k)
\mathbf{X}
= \sigma(k)\mathsfbi{B}(k)\mathbf{X},
\label{lsa_matrix}
\end{equation} 
where $\mathsfbi{A}$ and $\mathsfbi{B}$ are known matrices and $\mathbf{X}$ is the eigenvector of function values. We solve for all possible pairs of $\mathbf{X}$ and $\sigma$ with \texttt{eig}, the in-built eigenvalue solver in MATLAB.

The classic Orr-Sommerfeld problem is known for its non-normality---its eigenvalues are highly sensitive to perturbations \citep{Reddy1993}. To ensure that the eigenvalues converge, we eliminate rows in the matrix problem \eqref{lsa_matrix} that are independent of $\sigma$ \citep{Weideman2000}---this removes spurious modes which disrupts the convergence of the physical spectrum \citep{Goussis1989}. Furthermore, we precondition the problem by rescaling each row with $\|\mathbf{a}_j\|_1$, where $\mathbf{a}_j$ is the $j$-th row of $\mathsfbi{A}$ \citep{Wathen2015}. We note that without pre-conditioning, the numerical solutions of the eigenvalue problem can exhibit convergence issues between consecutive values of $N$ if $N$ is sufficiently large. 

We solve the eigenvalue problem with a starting value of $N=80$ Chebyshev nodes. We increase the number of nodes in intervals of $20$ and recalculate the spectrum until the most unstable eigenvalue is within $0.1\%$ of the previous estimate in the $L_2$--norm.

\subsection{Typical unstable modes of the eigenvalue problem}
\begin{figure}
	\centering
		\includegraphics[width=\textwidth]{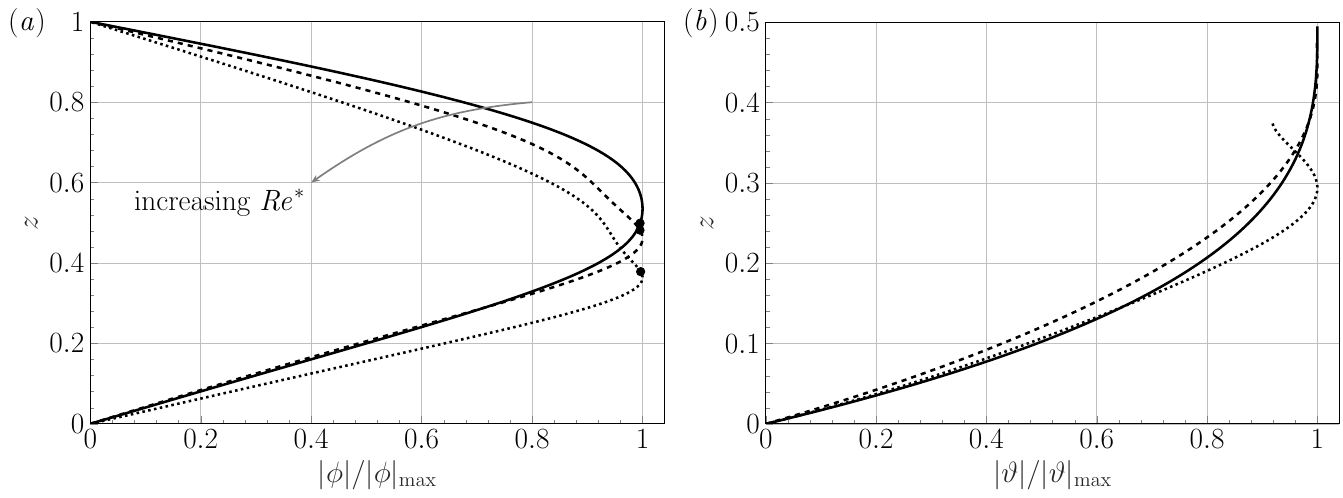}
	\caption{Magnitudes of unstable eigenmodes $\phi$ and $\vartheta$ as normalised functions of $z$ when $h=0.5$, $C_Y/\Rey^{*2}=10^{-3}$, $\lambda=1$, and $k=2$. Results are shown with $\Rey^* = 279.799556$ (solid), $985.757711$ (dashed), and $2663.736045$ (dotted). The solid circles on the stream functions in (\textit{a}) indicate their respective canopy height at steady state.}
	\label{fgr:bvp}
\end{figure}
\noindent Typical results for $\phi$ and $\vartheta$ are shown in \fig{bvp}, which correspond to flow perturbation and plant deflection respectively. 

We first note from the behaviour of $|\phi|$ that for all cases, the energy of the flow perturbation is localised near the top of the canopy. For slower flows, $|\vartheta|$ increases monotonically in $z$ so that the perturbations to the deflection angle are also largest at the top of the canopy. Moreover, for faster flows, the largest perturbation to the deflection may be in the middle of the canopy. The temporal evolution of such a mode is illustrated in \fig{monami}. 


\subsection{Temporal evolution of a perturbed steady configuration}

\noindent We can visualise how the physical configuration evolves in time by perturbing the base flow with a single unstable mode. We plot the streamlines of the flow given by
\begin{align}
\begin{pmatrix} u\\ w\end{pmatrix} =
\begin{pmatrix} u^*(z) \\ 0 \end{pmatrix} + \gamma e^{\mathrm{Re}(\sigma)t}
\Real\left[\begin{pmatrix}\phi'(z)\mathrm{e}^{\mathrm{i}[kx+\mathrm{Im}(\sigma)t]}\\
-\mathrm{i}k\phi(z)\mathrm{e}^{\mathrm{i}[kx+\mathrm{Im}(\sigma)t]}\end{pmatrix}\right],\label{perturbation}
\end{align}
where $\gamma$ is the (arbitrarily chosen) initial amplitude. 

\begin{figure}
	\centering
	\begin{overpic}[width=\textwidth]{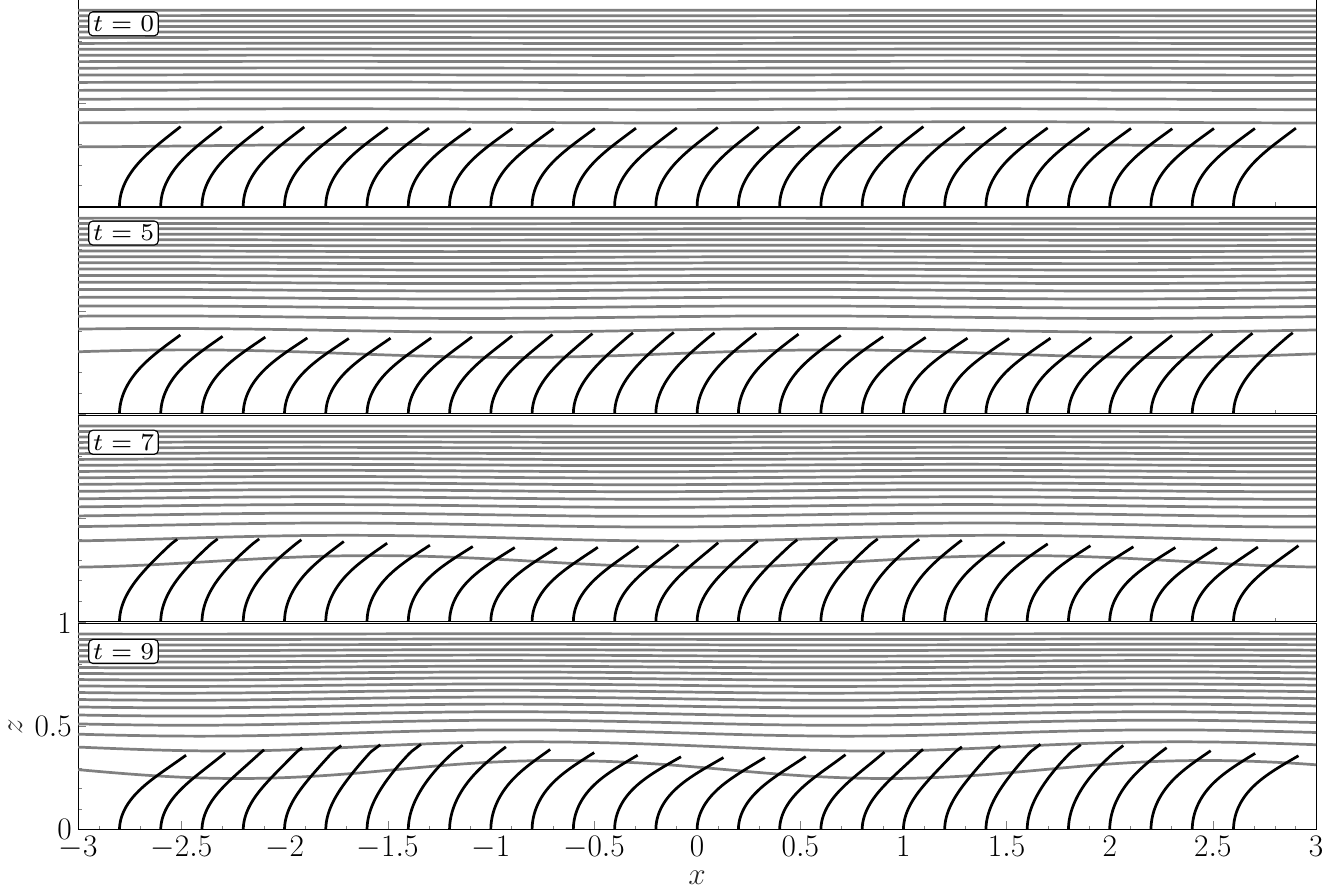}
	\end{overpic}
	\caption{Temporal evolution of a steady unidirectional base flow in the domain due to the growth of a single unstable eigenmode. The configuration of individual plants in the canopy are represented by black solid lines (to scale) along the bottom of the domain with streamlines in grey. For the eigenmode, $\phi$ and $\vartheta$ are rescaled by $|\phi|_{\mathrm{max}}$ before we impose an initial amplitude of $\gamma =10^{-3}$ in \eqref{perturbation}. Shown is the case where $h=0.5$, $\Rey^*=2663.736045$, $\lambda=1$, $C_Y/\Rey^{*2}=10^{-3}$ and $k=2$.}
	\label{fgr:monami}
\end{figure}

In \fig{monami}, we present the perturbed configuration at four different instances. The amplitude of the travelling wave grows in time as it convects downstream. If we plot the streamlines in \fig{monami} without the base flow, we will find closed contours along the top of the canopy corresponding to rolling vortices. Regarding the canopy configuration, plants oscillate synchronously, resembling monami. In particular, we see that when the deflection near the base of the canopy is increased, the canopy becomes more aligned with the flow, reducing the drag higher up the canopy, so that the deflection angle higher up is reduced. We highlight that our flexible canopy model is able to capture the streamwise variation of the canopy height and the local angle of deflection for each plant. 

In the remainder of the paper, we are interested in the following aspects: (i) in the limits of small and large canopy density, and small and large flexibility, what is the resultant behaviour of the eigenvalue problem and consequently the flow instabilities; (ii) once the numerical procedure in \S\ref{ssec:LSA_num} is applied at numerous values of $\Rey^*$ and $k$, we can form the neutral stability curve and thus examine the critical conditions in the parameter space for the onset of instability. 

We shall begin in \S\S\ref{ssec:stab_sparse} and \ref{ssec:stab_dense} with discussion of the asymptotic reductions in the limits of small and large canopy density.  

\subsection{Stability of flows for sparse canopies}
\label{ssec:stab_sparse}
\begin{figure}
	\centering
	\includegraphics[width=\textwidth]{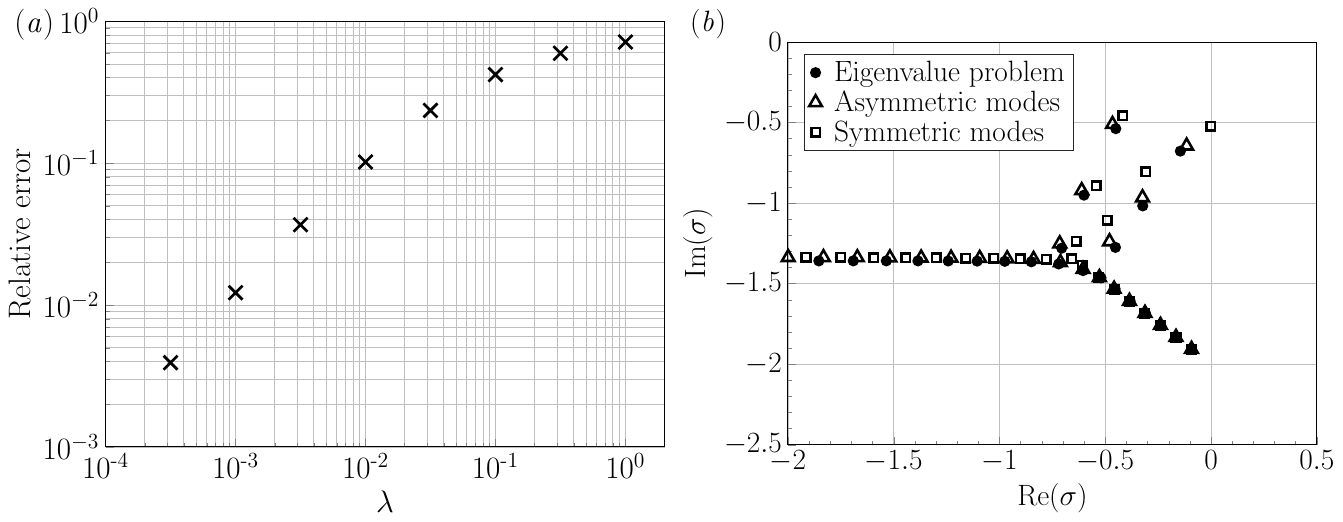}
	\caption{Comparison between the spectrum of the full eigenvalue problem \eqref{lsa_final} and some reduced problems. (\textit{a}) Relative error between the least stable eigenvalue of the rigid-canopy problem and the sparse canopy approximation  \eqref{sigma_uniform} for different values of $\lambda$. Shown is the case where $h=0.5$, $\Rey^*=200$ and $k=2$. (\textit{b}) Positions of the most unstable eigenvalues for the full problem and the corresponding symmetric and antisymmetric modes of the classic Orr-Sommerfeld problem \eqref{eq:orr_som}. Shown is the case where $h=0.5$, $\Rey^*=11610.6171$, $\lambda=10^{5}$, $C_Y/\Rey^{*2}=10^{-4}$ and $k=2$.}
	\label{fgr:lambda}
\end{figure}

\noindent In the limit where the canopy is sparse, in the sense that $\Rey^*\lambda,\,\Rey^{*2}\lambda\ll 1$, we know from \S3.2 that $u_0\sim 1$ and the momentum sink in \eqref{lsa_final1} becomes negligible. Since the perturbations of the beams are then decoupled from the fluid, we can approximate the eigenvalues of the full problem by solving
\begin{equation}
\frac{1}{\Rey^*}(\phi''''-2k^2\phi''+k^4\phi) =
(\sigma+\mathrm{i}k)(\phi''-k^2\phi),
\end{equation}
with
\begin{equation}
\phi(0)=0,~\phi(1)=0,~\phi''(0)=0,~\phi''(1)=0.
\label{os_bc}
\end{equation}
We can solve this reduced problem analytically and deduce that 
\begin{equation}
\sigma(k) =  -\frac{1}{\Rey^*}(k^2+n^2\upi^2) - k\mathrm{i},
\label{sigma_uniform}
\end{equation}
for $n\in\mathbb{N}$. The uniform flow is globally stable. We compare the leading eigenvalue with that of the rigid-canopy problem in \fig{lambda}a. 

In the distinguished limit where $\Rey^*\lambda\ll1$ with $\Rey^{*2}\lambda$ being $O(1)$, we can apply the approximations for the velocity profile in \S\ref{ssec:rigid_canopy} for rigid canopies and rewrite the eigenvalue problem as
\begin{equation}
\phi''''-2k^2\phi''+k^4\phi =
(\sigma_\mathrm{new}+\mathrm{i}k\Rey^{*2}\lambda\tilde{u})(\phi''-k^2\phi) - \mathrm{i}k\Rey^{*2}\lambda\tilde{u}''\phi+O(\Rey^*\lambda),
\label{orr_som_sparse}
\end{equation}
with $\phi$ satisfying the boundary conditions \eqref{os_bc}, $\tilde{u}$ given by \eqref{asy_sparse}, and $\sigma_\mathrm{new}=\Rey^*(\sigma+k\mathrm{i})$. This scaling determines the critical Reynolds number for instability when the canopy is sparse---we will verify this numerically in \S\ref{sec:6_crit}. 

\subsection{Stability of flows for dense canopies}
\label{ssec:stab_dense}
\noindent Recall from our dense-canopy analysis in \S\ref{subsec:dense_canopy} that in the limit $\lambda \to \infty$, the flow inside the canopy satisfies $u^* = O(\lambda^{-1/2})$ from \eqref{asy_dense_outer}. Moreover, the flow above the canopy is parabolic and given by \eqref{asy_dense_above} and \eqref{eq:pxseries}. Thus with this approximation, we should recover the eigenvalues of the classic Orr-Sommerfeld problem for a plane Poiseuille flow between $[h,2-h]$ with $u_0 = 0$ at $z = h$ (the channel bottom) and $u_0 = 1$ at $z = 1$ (the flow centreline). This gives [compare with \citealt{Drazin2004}]
\begin{subequations} \label{eq:orr_som}
\begin{equation}
\frac{1}{\Rey^*}(\phi''''-2k^2\phi''+k^4\phi) =
(\sigma+\mathrm{i}ku_0)(\phi''-k^2\phi)-\mathrm{i}ku_0''\phi,
\label{orr_som}
\end{equation}
with
\begin{equation}
\phi(h)=0,~\phi'(h)=0,~\phi(2-h)=0,~\phi'(2-h)=0.
\end{equation}
\end{subequations}
Recall from \citet{Chapman2002} that the above problem has two types of modes: symmetric and antisymmetric. Since we have imposed a no-shear condition \eqref{lsa_bc1} at $z=1$, eigenmodes of  the full problem \eqref{lsa_final} in this limit will only correspond to the antisymmetric modes of the classic problem \eqref{eq:orr_som}, as seen in \fig{lambda}b. These modes are always stable \citep{Orszag1971}. Therefore, our problem is also linearly stable in the dense-canopy limit.

\section{Critical conditions for the onset of instability}
\label{sec:6_crit}
We have thus seen that the system is linearly stable when the canopy is absent or infinitely dense. However, we found that unstable modes exist for intermediate canopy densities. Therefore, it is natural for us to investigate the critical conditions for the onset of instability. 

\subsection{Evolution of the neutral curve}
\label{ssec:evolution_neutral}
%
\begin{figure}
	\centering
	\includegraphics[width=\textwidth]{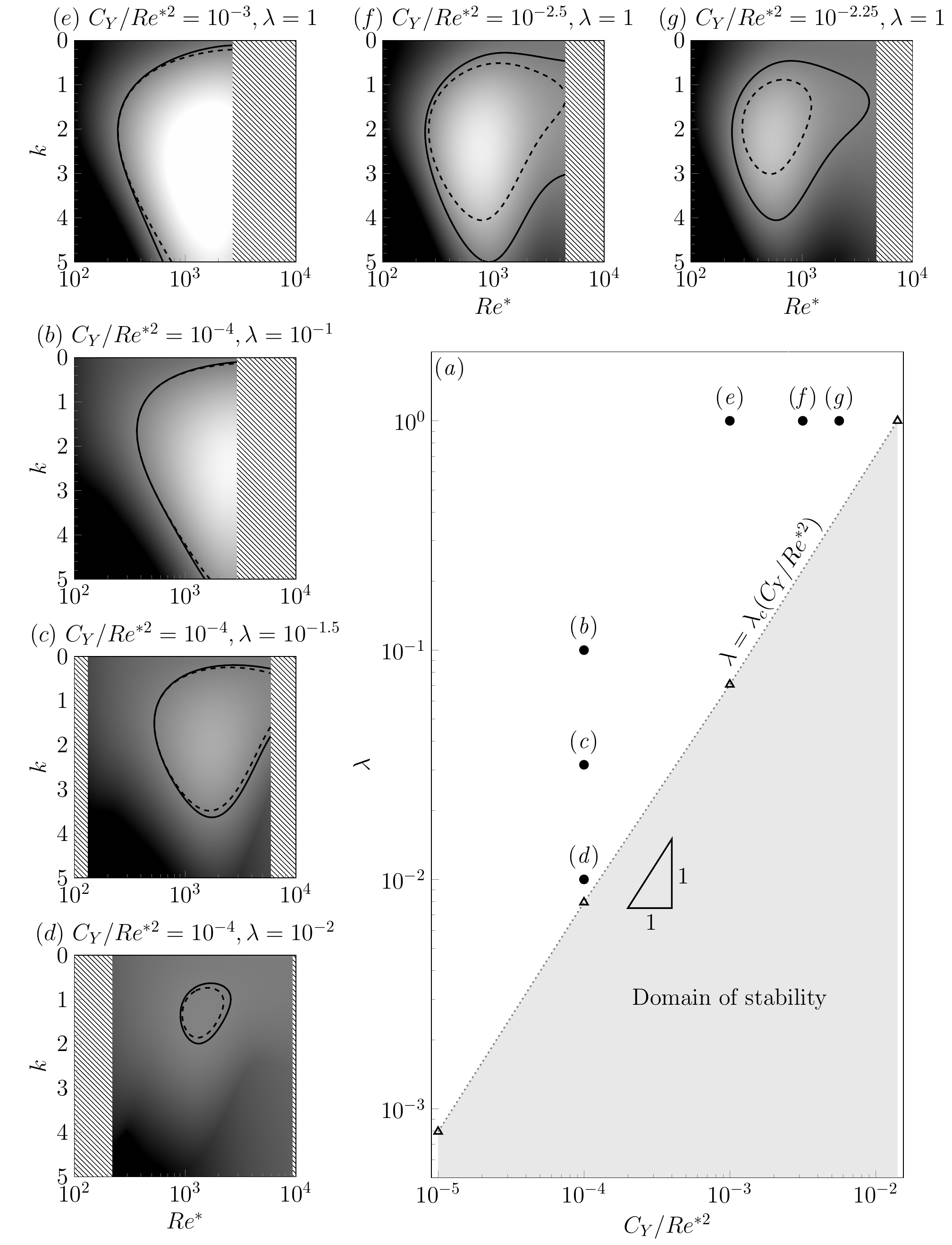}
	\caption{The main figure (a) shows the $(C_Y/Re^{*2}$, $\lambda)$-parameter space corresponding to elasticity and canopy density. The triangular markers ($\triangle$) are numerically determined solutions that lie on a critical curve for instability (sketched dotted). The six insets (\textit{b}) through (\textit{g}) show heat maps of the eigenvalue problem \eqref{lsa_final} in the $(\Rey^*, k)$-plane; colours correspond to values of $\Real(\sigma)$ that range from $-0.2$ (black) to $0.2$ (white); in each heat map, the neutral stability curve, $\Real(\sigma)=0$, is shown solid; areas that have not been swept are hatched. In \S\ref{ssec:ignore_canopy}, we present a decoupled model, and the neutral curve for that model is shown dashed within the heat maps.  
	}
	\label{fgr:neutral}
\end{figure}
\noindent We show in \fig{neutral} contour plots of $\Real(\sigma)$ for various points in the $(C_Y/Re^{*2}$, $\lambda)$-parameter space. Firstly, for a given canopy density and flexibility, the base states first become unstable for a critical wavenumber $k$ of $O(1)$. Secondly, although not plotted here, all of the unstable modes have $\Imag(\sigma)<0$, with $|\Imag(\sigma)|$ increasing non-linearly with $k$ for any given $\Rey^*$, corresponding to the instabilities being convective downstream and dispersive. This is in agreement with the predictions by \citep{Singh2016,Luminari2016} on flows through rigid canopies. 

We comment now on a key feature of \fig{neutral} that will be further discussed in \S\S\ref{ssec:crit_re} and \ref{ssec:ignore_canopy}. Notice that for a fixed flexibility, $C_Y/\Rey^{*2}$, and decreasing density, $\lambda$, \textit{i.e.} as we transition from (\textit{b}) to (\textit{c}) to (\textit{d}), the topology of the neutral curve may change: the area enclosed by the neutral curve tends to zero and the system becomes globally stable. This trend also occurs for fixed $\lambda$ and for increasing $C_Y/\Rey^{*2}$ \textit{i.e.} as we transition from (\textit{e}) to (\textit{f}) to (\textit{g}). Thus, there exists a critical curve, 
\begin{equation}\label{lambda_crit}
	\lambda = \lambda_c(C_Y/\Rey^{*2}),
\end{equation}
in the $(C_Y/\Rey^{*2}, \lambda)$-plane that separates the two possibilities---globally stable for all $\Rey^*$, or unstable for some $\Rey^*$. This bifurcation is plotted as a line in \fig{neutral}. In particular, our numerical results suggest that $\lambda_c\sim C_Y/\Rey^{*2}$. 

\subsection{Behaviour of the critical Reynolds number}
\label{ssec:crit_re}
\begin{figure}
	\centering
	\includegraphics[width=\textwidth]{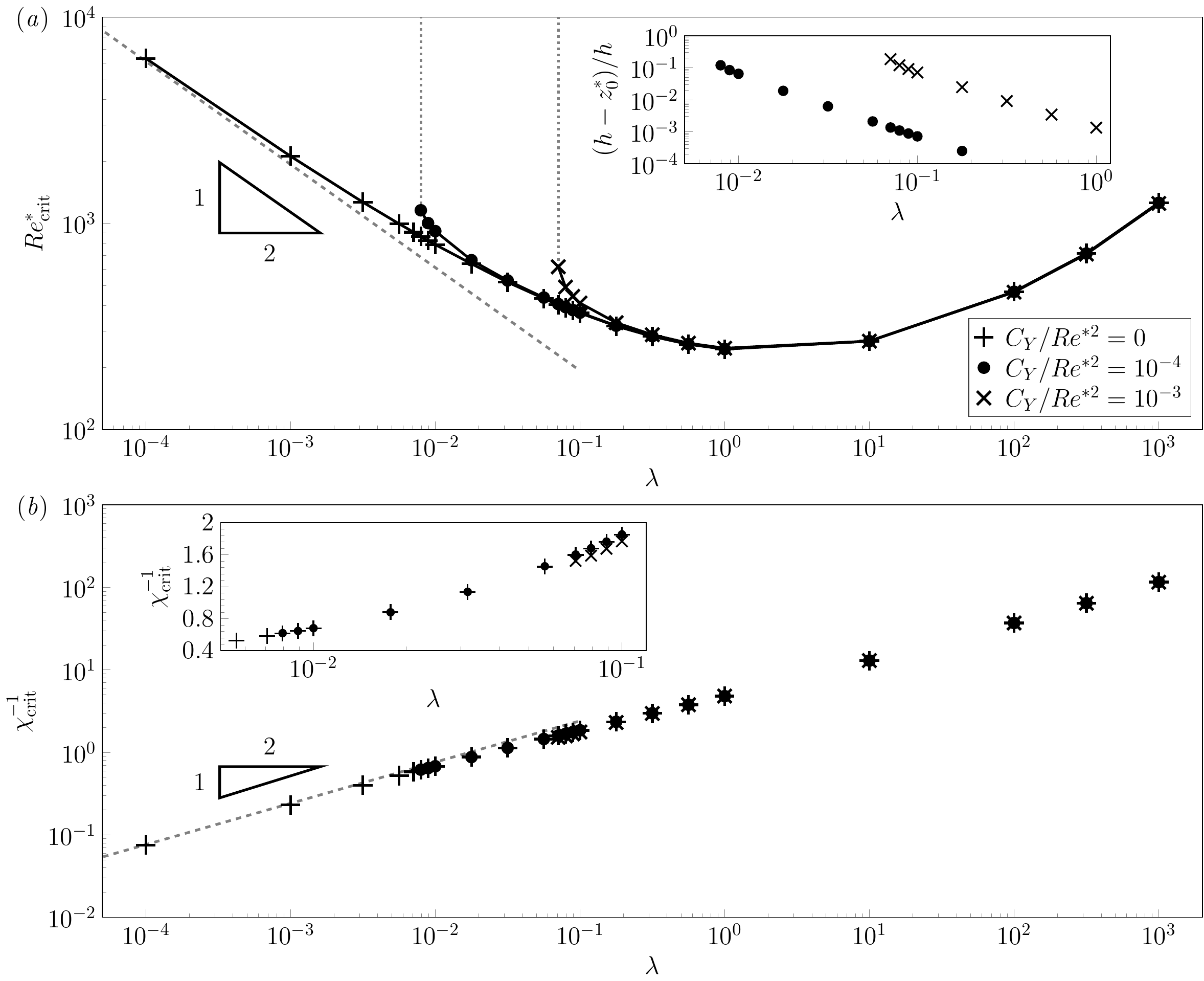}
	\caption{(\textit{a}) Critical Reynolds number, $\Rey_{\mathrm{crit}}^{*}$, and (\textit{b}) critical shear at the top of the canopy, $\chi_{\mathrm{crit}}^{-1}$, for flows with different canopy densities. Shown is the case where $h=0.5$. Symbols are used to indicate the flexibility of the plants, as stated in the legend of (\textit{a}). The lines which connect the data points act as visual guides with vertical dotted lines indicating  that the system is stable for $\lambda<\lambda_c$ \eqref{lambda_crit}. The grey dashed line indicate the scaling prediction \eqref{orr_som_sparse} for rigid canopies, by noting that $\chi\sim (\Rey^*\lambda)^{-1}$ in the limit $\Rey^*\lambda\ll 1$ \eqref{z_u}. The inset in (\textit{a}) gives the relative vertical displacement of the canopy at the corresponding steady state and the inset in (\textit{b}) highlights the difference in the critical shear for sparse canopies. }
	\label{fgr:crit_lambda}
\end{figure}
\noindent We notice from the contour plots in \fig{neutral} that if there are any unstable modes in the parameter space, then for every neutral curve, there exists a critical value for $\Rey^*$, $\Rey_{\mathrm{crit}}^*>0$, such that flows will be first unstable if $\Rey^*>\Rey_{\mathrm{crit}}^*$. Therefore, it is natural for us to analyse the behaviour of $\Rey_{\mathrm{crit}}^*$ as a function of $\lambda$ and $C_Y$, shown in \fig{crit_lambda}. 

For fixed flexiblity, $C_Y/\Rey^{*2}$, we see that the critical Reynolds number, $\Rey_{\mathrm{crit}}^*$, exhibits a minimum as the density, $\lambda$, changes. This can be understood intuitively through the asymptotic analysis of \S\ref{ssec:stab_sparse} and \S\ref{ssec:stab_dense}. In particular, as the canopy density decreases to zero or increases to infinity, the base flow becomes globally stable. We will revisit this statement when we analyse the critical shear in \S\ref{ssec:crit_shear}.

The critical values $\lambda=\lambda_c(C_Y/\Rey^{*2})$ below which the flow is globally stable are indicated by the vertical dotted lines in \fig{crit_lambda}.

\subsection{Behaviour of the critical shear}
\label{ssec:crit_shear}
\noindent In the previous discussion, we have sought to understand the effects of plant flexibity and density on the critical value of $\Rey^*$ separating unstable and stable flows. However, there are interesting insights once this is viewed in regards to the shear. 

We examine the critical shear in our model for the onset of instability. Suppose we let $\chi$ be the (dimensionless) Navier slip length at steady state at $z=z_0^*$, namely
\begin{equation}
\chi= \frac{u^*(z_0^*)}{{u^*}'(z_0^*)}.
\label{slip}
\end{equation} 
For every flow that is marginally stable, we can then quantify the critical shear with $\chi_{\mathrm{crit}}^{-1}$, the reciprocal of the Navier slip length. 

We observe from \fig{crit_lambda}b that for any given canopy density:
\begin{enumerate}[label=(\roman*),leftmargin=*, align = left, labelsep=\parindent, topsep=3pt, itemsep=2pt,itemindent=0pt ]
	\item The critical shear, $\chi_{\mathrm{crit}}^{-1}$, increases monotonically with $\lambda$.
	\item  The critical shear, $\chi_{\mathrm{crit}}^{-1}$, is nearly identical across different plant flexibilities (cf. the inset of \fig{crit_lambda}b) even when $\Rey_{\mathrm{crit}}^{*}$ and the configuration of the canopies are distinct. This is highlighted by the inset of \fig{crit_lambda}a.
\end{enumerate}
The collapsed data suggests that shear at the top of the canopy is the relevant criterion in determining the stability of steady unidirectional flows. This conclusion is consistent with previous statistical analysis on turbulent flows through vegetation, which suggested that canopy-scale instabilities are shear induced \citep{Raupach1996, Finnigan2000}. 

For sparse rigid canopies, we found that stability depended on $\Rey^{*2}\lambda$, implying  $\Rey_{\mathrm{crit}}^*\sim\lambda^{-1/2}$ as $\lambda\rightarrow 0$. Since $\chi\sim(\Rey^*\lambda)^{-1}$, this implies $\chi_{\mathrm{crit}}\sim \lambda^{-1/2}$ as $\lambda\rightarrow 0$. The scaling laws that we predicted in \S\ref{ssec:stab_sparse} for the limit $\Rey^*\lambda\ll1$ with $\Rey^{*2}\lambda$ being $O(1)$ are indicated by the dashed lines in \fig{crit_lambda}. 

Finally, we interpret the U-shaped curve in \fig{crit_lambda}a physically. Denoting by $\Lambda$ the canopy density corresponding to the minimum $\Rey_{\mathrm{crit}}^*$, we find that for canopies with $\lambda > \Lambda$, the shear threshold is greater and therefore a faster flow is required for instability. For $\lambda < \Lambda$, the shear threshold is lower but because of the reduced drag, a faster flow is required to generate this shear at the top of the canopy. 

\subsection{What happens if the perturbation of the canopy is ignored?} 
\label{ssec:ignore_canopy}
\noindent Despite the complexity of the solutions presented above, it appears that the overall qualitative behaviour of the neutral stability curves shown in \fig{neutral} can be understood through largely hydrodynamical effects. In particular, we may attempt to decouple the canopy and the flow by considering only perturbations in the flow. This is equivalent to setting the perturbations on the canopy, $\vartheta$, and stresses, $\mathcal{T}_{1,3}$, to zero in \eqref{lsa_wave} and solving the single equation \eqref{lsa_final1} for $\phi$. Once this is done, the neutral stability curves for this decoupled problem are derived, and these are shown in \fig{neutral} as dashed lines. 

When the canopy is dense or the plants are stiff, the difference between the two problems are negligible, as discussed in \S\ref{ssec:flex_canopy}. However, as we decrease the canopy density or increase the plant flexibility, the modes of the fluid problem are always more stable than the coupled problem, indicating that the movement of the canopy contributes towards instability. 

\section{Conclusions}
\label{sec:7_con}

We have shown that the mechanics and instabilities of flows over a fully submerged vegetative region can be studied using relatively simple models. In particular, we have presented a continuum approximation that effectively homogenises the vegetation, and identifies five key dimensionless parameters: the Reynolds number, Froude number, submergence ratio, canopy density, and the Cauchy number. The study of two-dimensional steady unidirectional flows allows for the derivation of a number of interesting results, and asymptotic analysis can be used to study particular cases of rigid or flexible canopies, and dense or sparse vegetation.  

Temporal stability of steady unidirectional flows can be studied by numerically solving an eigenvalue problem for perturbations in the form of travelling waves. In both limits where the canopy density, $\lambda\rightarrow 0$ and $\lambda \rightarrow \infty$, we have demonstrated that the base configurations are stable. However, for intermediate canopy densities, there exists a critical Reynolds number and critical shear (determined by the flow at the top of the canopy) that separates unstable and stable flows. These critical thresholds vary as functions of the canopy density and flexibility. In particular, we have shown that plant deflection encourages the temporal growth of perturbations. Our flexible canopy model can also capture the temporal evolution of monami by studying the growth of a single unstable eigenmode. In contrast to typical terrestrial flows, the local angle of deflection may not increase monotonically in height.

\section{Discussion}
\label{sec:8_discuss}


One important conclusion from the present study is that the shear at the top of the canopy is a dominant factor in determining the stability of a flow; moreover, we have shown how this result can be ascertained using relatively compact models of vegetative flows. Thus one interesting question that follows is whether there may be particular flow regimes where more involved models are required, and for which the central mechanism for instability may be different. As a particular example, we highlight the review by \citet{Nepf2012}, who notes that as the canopy becomes increasingly sparse, the flow transitions from a mixing-layer-like flow to a boundary-layer flow (see \emph{e.g.} \fig{crit_lambda} as $\lambda \to 0$). In our current model we have neglected the effects of bed shear, and this hints at the need for a model for which the sparse-canopy limit can be more accurately captured. 

Another important avenue for progress is the consideration of the particular turbulence model used, and again, there are signs that a more complete model is required. Note that we analysed flow stability using a single mixing length at the top of the canopy \eqref{slip} and our constant eddy viscosity \eqref{eddy} closure model can be interpreted as effectively averaging the various length scales involved. However, as we have highlighted in the asymptotic analysis of \S\ref{sec:3_uni}, in limits of small or large elasticity/density, it is unclear whether it may be necessary to consider a variable eddy viscosity in order to capture disparate length scales in the flow.

\emph{Which length scales might be involved in a more complete model?} Firstly, the bed can be rough in reality and the boundary layer along the bed may contribute one length scale. Secondly, there is an element length scale inside the canopy, which is determined by the wakes of individual plants. Finally, for deeply submerged canopies ($h\ll 1$), turbulence that is generated far below is expected to be negligible far above ($z/h\gtrapprox 2$); thus there is a length scale determined by the decay of the canopy-scale vortices. A more detailed description of each scale can be found in \citet{Marion2014}, and we envision that a more refined model will impose a closure model that captures a number of these length scales. However, in order to implement such a scheme, we require a better understanding of the importance of these length scales, and moreover, we would hope to determine these scales as part of the model, rather than verified \textit{a posteriori} from given parameters [see discussion in \citet{Poggi2004a}]. 

In addition to the typical vegetative fluid-structure questions we have highlighted above, there is a great deal of interest as well in the study of intrinsically time-dependent scenarios which include the case of oscillatory flows through vegetation. Compared to unidirectional flows of the same magnitude, oscillatory flows can have higher in-canopy flow velocities \citep{Lowe2005} and they can potentially create more shear than flows over a bare bed \citep{Zhang2018}. For these situations, vegetation can move out of phase with the waves \citep{GijonMancheno2016}. Moreover, drag can be affected by the wave frequency \citep{Mattis2019} and the presence of a current \citep{Zeller2014,Luhar2016,Lei2019}. A natural extension of our work is to utilise our coupled model in \S\ref{sec:2_model} to analyse the interaction between oscillatory flows and vegetation. 

\mbox{}\par
\noindent\textbf{Acknowledgement}

This publication is based on work partially supported by the EPSRC Centre For Doctoral Training in Industrially Focused Mathematical Modelling (EP/L015803/1) in collaboration with HR Wallingford and US Army Engineer Research and Development Center. We would like to thank A. Dimakopoulos
from HR Wallingford and C. Kees from US Army ERDC for their important contributions. We are also grateful to I. Hewitt and D. Vella for insightful discussions.

\appendix
\section{Jump conditions for the Orr-Sommerfeld equation}\label{app:A}
In this appendix, we discuss the derivation of the jump conditions of $\phi$ at $z=z_0^*$ \eqref{lsa_phi''}--\eqref{lsa_phi'''} presented in \S\ref{ssec:travel}. By equating the derivatives of $\hat{p}$ in \eqref{lsa_first2}--\eqref{lsa_first3}, and consider travelling wave perturbations \eqref{lsa_wave}, $\phi$ satisfies 
\begin{eqnarray}
\hspace{-0.5cm}\frac{\phi''''-2k^2\phi''+k^4\phi}{\Rey^*} &=&(\sigma+\mathrm{i}ku^*)(\phi''-k^2\phi)+\mathrm{i}k{u^*}''\phi +\mathcal{H}_\mathrm{c}'+\mathrm{i}k\mathcal{H}_\mathrm{s}+\mathcal{J}_\mathrm{c}'+\mathrm{i}k\mathcal{J}_\mathrm{s},
\label{lsa_final1_appendix}
\end{eqnarray}
with
\begin{align}
\mathcal{H}_c &= \lambda\mathrm{H}(z_0^*-z)u^*\cos\theta^*\left[\cos\theta^*\left(\phi'-\sigma\mathcal{A}\right)-\sin\theta^*\left(u^*\vartheta-ik\phi+\sigma\mathcal{B}\right)\right],\\
\hspace{-2cm} \mathcal{H}_s& = \lambda\mathrm{H}(z_0^*-z)\left[\frac{1}{2}u^{*2}\vartheta+u_0\sin\theta^*\left[\cos\theta^*\left(\phi'-\sigma\mathcal{A}\right)-\sin\theta^*\left(u^*\vartheta-ik\phi+\sigma\mathcal{B}\right)\right]\right],\\
\mathcal{J}_c & = -\frac{\lambda}{2}\delta(z-z_0^*)u^{*2}\cos^2\theta^*\mathcal{C}(z_0^*)\cos\theta(z_0^*),\\
\mathcal{J}_s&= -\frac{\lambda}{2}\delta(z-z_0^*)u^{*2}\sin\theta^*\cos\theta^*\mathcal{C}(z_0^*)\cos\theta(z_0^*),
\end{align}
The expressions for $\mathcal{A}$, $\mathcal{B}$, and $\mathcal{C}$ are given by \eqref{lsa_C}--\eqref{lsa_B} in the main text. The modified Orr-Sommerfeld equation \eqref{lsa_final1_appendix} here can be rewritten into equation \eqref{lsa_final1} by splitting the equation into cases above and below $z=z_0^*$. The corresponding jump conditions can be derived from integrating \eqref{lsa_final1_appendix} across the interval $[z_0^*-d,z_0^*+d]$, with $d\rightarrow 0$. 



\end{document}